\documentclass[tradiabstract]{aa} % for the abstract without structuration
\usepackage{graphicx}
\usepackage{txfonts}
\usepackage{natbib}
\usepackage{longtable}
\usepackage{lscape}
\usepackage{multirow}

\bibpunct{(}{)}{;}{a}{}{,}

\newcommand{\kms}{km\,s$^{-1}$}
\newcommand{\ms}{m\,s$^{-1}$}

\newcommand{\mum}{$\mu$m}

\newcommand{\Msun}{$M_{\odot}$}
\newcommand{\Mearth}{$M_{\oplus}$}
\newcommand{\Lsun}{$L_{\odot}$}

\newcommand{\nodata}{...}
\newcommand{\accunit}{$M_{\odot}$\,yr$^{-1}$}
\newcommand{\rev}{ }
\newcommand{\newrev}{ }

\newcommand{\subrev}{ }

\newcommand{\fig}{Fig.}
\newcommand{\sect}{Sect.}

\begin{document}
%
%\title{Disk properties in the nearby and young $\epsilon$\,Cha Association\thanks{Based on observations performed at ESO's La Silla-Paranal observatory under programme 076.C-0470}-- Evolution in the sparse associations}

\title{GW~Ori: Inner disk readjustments in a triple system}

\author{M.~Fang\inst{1} \and A.~Sicilia-Aguilar\inst{2, 1} \and V.~Roccatagliata\inst{3} \and D.~Fedele\inst{4} \and Th. Henning\inst{5} \and C.~Eiroa\inst{1} \and A. M\"uller\inst{6}}
\institute{Departamento de F\'{\i}sica Te\'{o}rica, Universidad Aut\'{o}noma de Madrid, Cantoblanco 28049, Madrid, Spain \and SUPA, School of Physics and Astronomy, University of St Andrews, North Haugh, St Andrews KY16 9SS, UK \and Universit\"ats-Sternwarte M\"unchen, Ludwig-Maximilians-Universit\"at, Scheinerstr. 1, 81679, M\"unchen, Germany \and Max Planck Institut f\"ur Extraterrestrische Physik, Giessenbachstrasse 1, 85748, Garching, Germany \and Max-Planck-Institut f\"ur Astronomie, K\"onigstuhl 17, 69117 Heidelberg, Germany \and European Southern Observatory, Alonso de Cordova 3107, Vitacura, Santiago, Chile}
 \date{Received 6 May 2014 \/ accepted 7 July 2014}
 \abstract{Disks are expected to dissipate quickly in  binary or multiple systems. Investigating such systems can improve our knowledge of the disk dispersal. The triple system GW~Ori, still harboring a massive disk, is an excellent target.}
%Aims:
{We study the young stellar system GW~Ori, concentrating on its accretion/wind activity and disk properties.}
%Methods:
{We use high-resolution optical spectra of GW~Ori to do spectral classification and derive the radial velocities (RV). We analyze the wind and accretion activity using the emission lines in the spectra. We also use $U$-band photometry, which has been collected from the literature, to study the accretion variability of GW~Ori. We characterize the disk properties of GW~Ori by modeling its spectral energy distribution (SED).}
%Results:
{{\rev By comparing our data to the synthetical spectra, we classify GW~Ori as a G8 star}. Based on the RVs derived from the optical spectra, we confirm the previous result as a close companion in GW~Ori with a period of $\sim$242~days and an orbital semi-major axis of $\sim$1\,AU. The RV residuals after the subtraction of the orbital solution with the equivalent widths (EW) of accretion-related emission lines vary with periods of 5--6.7\,days during short time intervals, which are caused by the rotational modulation. The H$\alpha$ and  H$\beta$ line profiles of GW~Ori can be decomposed in two central-peaked emission components and one blue-shifted absorption component. The blue-shifted absorption components are due to a disk wind modulated by the orbital motion of the close companion. Therefore, the systems like GW~Ori can be used to study the extent of disk winds. We find that the accretion rates of GW~Ori are rather constant but can occasionally be enhanced by a factor of 2--3. We reproduce the SED of GW~Ori by using disk models with gaps $\sim$25--55\,AU in size. A small population of tiny dust particles within the gap  produces the excess emission at near-infrared bands and the strong and sharp silicate feature at 10\,\mum. The SED of GW~Ori exhibits dramatic changes on timescales of $\sim$20\,yr in the near-infrared bands, which can be explained as the change in the amount and distribution of small dust grains in the gap. We collect a sample of binary/multiple systems with disks in the literature and find a strong positive correlation between their gap sizes and separations from the primaries to companions, which is generally consistent with the prediction from the theory.}
{}

 \maketitle

\section{Introduction}
Young stars are born with disks as a result of angular momentum conservation \citep{1977ApJ...214..488S}.  Observations suggest disk lifetimes of a few Myrs \citep{1989AJ.....97.1451S,2001ApJ...553L.153H, 2002astro.ph.10520H,2007ApJ...662.1067H,2006ApJ...638..897S,2010A&A...510A..72F,2012A&A...539A.119F,2013A&A...549A..15F}. However, the physical processes in the disk evolution are still poorly understood. {\newrev Spectroscopic and/or imaging surveys toward nearby field dwarfs suggest that 50\% of G-type stars have companions \citep{1991A&A...248..485D}, and 30\%--40\% of M-type stars are in binary/multiple systems \citep{1992ApJ...396..178F,2012ApJ...754...44J}.} The fractions of binary/multiple systems  are even higher in star-forming regions \citep{1993AJ....106.2005G,1993A&A...278..129L,1997ApJ...481..378G,2008ApJ...683..844L,2011ApJ...731....8K}. Therefore, the interaction between disks and companions has been proposed as an efficient mechanism to dissipate disks \citep{1993prpl.conf..749L}. Observationally, this mechanism can  be very efficient in disk disperal at very early stages \citep[$<$1\,Myr,][]{2009ApJ...696L..84C,2012ApJ...745...19K} and may play a key role in dissipating the disks in sparse stellar associations \citep{2006ApJ...653L..57B,2013A&A...549A..15F}.

\begin{figure*}
\centering
\includegraphics[width=1.8\columnwidth]{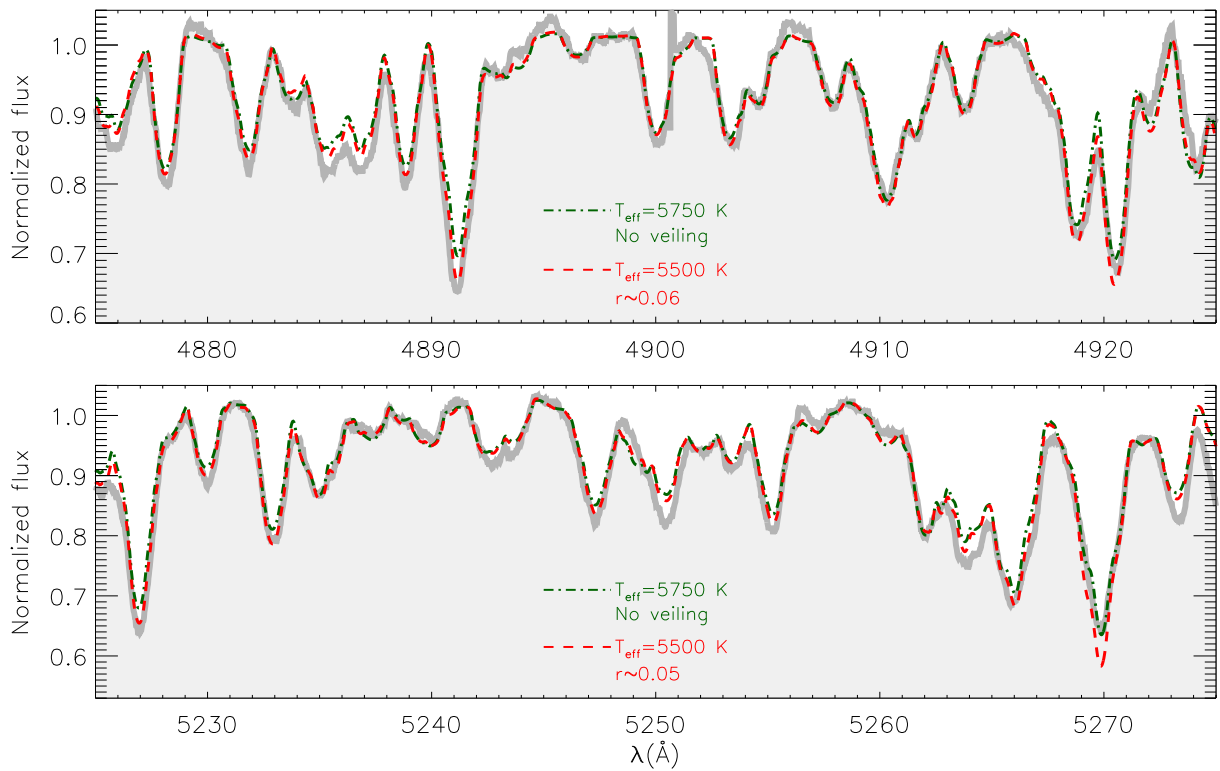}
\caption{Comparison of the observed spectrum (thick gray lines) of GW~Ori, which is combined by using the data observed during Jan. 6-11 in 2009, and synthetical spectra at $T_{\rm eff}$=5750 (the blue dash-dotted lines) and 5500\,K (the dashed lines). The spectra are all normalized. The best veiling value ($r$) is also given for each model spectrum.}\label{Fig:comparison_obs_synthetic}
\end{figure*}

In a young binary system, two types of disks could be present: a circumstellar disk  surrounding each star in the system and a circumbinary disk around the binary pair \citep{2000prpl.conf..731L}. Circumstellar disks can be truncated outside by the tidal companion-disk interaction and can be misaligned, while circumbinary disks can be carved out inside by the companions \citep{1993prpl.conf..749L,1994ApJ...421..651A,2011A&A...534A..33R}. Gaps, quickly produced in the inner regions of circumbinary disks, can separate circumstellar disks and circumbinary disks \citep{1993prpl.conf..749L,2000prpl.conf..731L}. Such gaps in circumbinary disks, as suggested in the  spectral energy distributions (SED) of some binary systems \citep{1997AJ....114..301J}, have been directly detected with millimeter interferometry \citep{2013ApJ...775..136R,2013ApJ...775...30I}.

For a circumbinary disk with a disk thickness-to-radius ratio h/r$>$0.05, the simulations indicate that the gap in the inner disk region can be replenished with material from the circumbinary disk in the form of gas streams, which can supply the mass for accretion onto the central binary \citep{1996ApJ...467L..77A}. The simulations also find that the periodic perturbations in the circumstellar disk caused by the orbital motion of the close companion can induce mass flow across the gap, resulting in accretion that changes with the orbital motion of binary. For a binary with a high eccentricity ($e$=0.5) and a mass ratio near one, the  accretion rate of the system can be strongly modulated in time and reach a maximum near periastron. However, when the eccentricity of a binary is lower ($e$=0.1),  the accretion rate of the system can be still pulsed with the orbital motions, but the enhancement can be smooth and less notable than those of the high-$e$ cases \citep{1996ApJ...467L..77A}. The scenario of pulsed accretion in binary systems, as predicted by  the simulations, is confirmed by the observations of only few cases, such as DQ~Tau and UZ~Tau~E \citep{1997AJ....114..781B,2005A&A...429..939M,2007AJ....134..241J}.  An investigation of other binaries can contribute in understanding the accretion processes in such systems.

The object GW~Ori, located at $\lambda$~Ori \citep[$\sim$400\,pc,][]{1977MNRAS.181..657M}, was first revealed as a spectroscopic binary (GW~Ori~A/B) with an orbital period of $\sim$242\,days and a separation of $\sim$1\,AU \citep{1991AJ....101.2184M}. The new near-infrared interferometric observations confirm the existence of a close companion (GW~Ori~B) and detect a second (GW~Ori~C) with a projected separation of $\sim$8\,AU from GW~Ori~A \citep{2011A&A...529L...1B}. Though it is a triple system, observations show that GW~Ori is still harboring a massive disk ($\sim$0.3\,\Msun) with a high accretion rate of $\sim$3$\times$10$^{-7}$\,\accunit \citep{1995AJ....109.2655M,2004AJ....128.1294C}. Thus, GW~Ori is a very interesting target and deserves further detailed investigation. 

\begin{table*}
\caption{Observing log for the spectroscopy.\label{Tab:obs_log}}
\centering
\begin{tabular}{ccccc|ccccc}
\hline\hline
ID &Obs. Date & JD-2450000 &Instrument & Exp. time (s) &ID &Obs. Date & JD-2450000 &Instrument & Exp. time (s)\\
\hline
1&2007-11-08& 4412.689&FEROS   &900.0&30&2009-01-13& 4844.538&FEROS   &960.0\\
2&2007-11-09& 4413.727&FEROS   &1300.0&31&2009-01-13& 4844.680&FEROS   &960.0\\
3&2008-01-31& 4496.530&FEROS   &1200.0&32&2009-01-14& 4845.606&FEROS   &960.0\\
4&2008-11-07& 4777.866&FEROS   &960.0&33&2009-02-12& 4874.520&HARPS   &540.0\\
5&2008-11-09& 4779.777&FEROS   &900.0&34&2009-02-13& 4875.532&HARPS   &540.0\\
6&2008-11-09& 4779.788&FEROS   &900.0&35&2009-02-14& 4876.574&HARPS   &540.0\\
7&2008-11-11& 4781.869&FEROS   &900.0&36&2009-02-15& 4877.543&HARPS   &540.0\\
8&2008-11-12& 4782.849&FEROS   &900.0&37&2009-03-01& 4891.508&HARPS   &600.0\\
9&2008-11-13& 4783.851&FEROS   &720.0&38&2009-03-01& 4892.501&HARPS   &600.0\\
10&2008-11-14& 4784.690&FEROS   &720.0&39&2009-03-03& 4893.519&HARPS   &600.0\\
11&2008-11-16& 4786.777&HARPS   &1200.0&40&2009-03-04& 4894.565&HARPS   &600.0\\
12&2008-11-17& 4787.747&HARPS   &1200.0&41&2009-04-28& 4950.469&FEROS   &960.0\\
13&2008-11-18& 4788.813&HARPS   &900.0&42&2009-10-01& 5105.810&FEROS   &465.8\\
14&2009-01-05& 4836.533&FEROS   &960.0&43&2009-10-02& 5106.849&FEROS   &960.0\\
15&2009-01-05& 4836.629&FEROS   &960.0&44&2009-10-04& 5108.867&FEROS   &1200.0\\
16&2009-01-06& 4837.538&FEROS   &960.0&45&2009-10-07& 5111.876&FEROS   &960.0\\
17&2009-01-06& 4837.599&FEROS   &960.0&46&2009-12-14& 5179.775&FEROS   &900.0\\
18&2009-01-07& 4838.538&FEROS   &960.0&47&2009-12-15& 5180.777&FEROS   &900.0\\
19&2009-01-07& 4838.689&FEROS   &960.0&48&2009-12-16& 5181.762&FEROS   &900.0\\
20&2009-01-08& 4839.567&FEROS   &960.0&49&2010-01-03& 5199.705&FEROS   &960.0\\
21&2009-01-08& 4839.720&FEROS   &960.0&50&2010-01-06& 5202.702&FEROS   &960.0\\
22&2009-01-09& 4840.560&FEROS   &960.0&51&2010-01-28& 5224.650&FEROS   &465.8\\
23&2009-01-09& 4840.687&FEROS   &960.0&52&2010-01-28& 5224.663&FEROS   &1200.0\\
24&2009-01-10& 4841.542&FEROS   &960.0&53&2010-01-29& 5225.664&FEROS   &900.0\\
25&2009-01-10& 4841.632&FEROS   &960.0&54&2010-01-30& 5226.644&FEROS   &900.0\\
26&2009-01-11& 4842.572&FEROS   &960.0&55&2010-02-27& 5254.566&FEROS   &900.0\\
27&2009-01-11& 4842.633&FEROS   &960.0&56&2010-03-05& 5260.595&FEROS   &900.0\\
28&2009-01-12& 4843.538&FEROS   &960.0&57&2010-03-13& 5268.535&FEROS   &900.0\\
29&2009-01-12& 4843.619&FEROS   &960.0&58&2010-03-14& 5269.541&FEROS   &900.0\\
\hline
\end{tabular}
\end{table*}

We obtain 58 high-resolution optical spectra of GW~Ori. As a complement, we also collect a large set of archive data, including a set of multi-epoch broad band photometry in different bands, and the infrared spectrum from the Spitzer InfraRed Spectrograph \citep[IRS,][]{2004ApJS..154...18H}. Using the accretion-related emission lines in the optical spectra and $U$-band photometry,  we investigate the accretion behavior of GW~Ori, which concentrates mainly on the accretion variability that has not been studied in detail yet. With the infrared data, we will characterize the disk properties of GW~Ori. We arrange this paper as follows. In \sect~\ref{Sec:data}, we describe the observations and reduction of the optical data. In \sect~\ref{Sec:result}, we present our results which are then discussed in \sect~\ref{Sec:discussion}. We summarize our results in \sect~\ref{Sec:summary}.

\section{Observations and data reduction}\label{Sec:data}
Our optical spectra of GW~Ori were taken with the Fiber-fed Extended Range Optical Spectrograph \citep[FEROS,][]{1999Msngr..95....8K}, which is mounted on the 2.2~m MPG/ESO (Max-Planck Gesellschaft/European Southern Observatory) telescope, and the High Accuracy Radial velocity Planet Searcher \citep[HARPS,][]{2003Msngr.114...20M}, which is mounted on the 3.6~m telescope. Both telescopes are located at La~Silla Observatory. The FEROS has a spectral resolution of $\lambda/\Delta\lambda$$\sim$48000 with a wavelength coverage of 3600--9200\AA. The HARPS has a higher spectral resolution ($\lambda/\Delta\lambda$$\sim$115000), but a narrower wavelength coverage (3800--6900\AA). With the two instruments, a total of 58 spectra were obtained during 2007--2010 (see Table~\ref{Tab:obs_log} for a detailed description of these observations). Both FEROS and HARPS have their own online data reduction pipelines, which can automatically produce science-quality spectra with calibrated wavelengths from observational raw data. 
Recently, \citet{2013A&A...556A...3M} found that the barycentric correction of the FEROS data reduction pipeline is inaccurate, as it induces an artificial one-year period with a semi-amplitude of 62\,\ms. Following \citet{2013A&A...556A...3M}, we apply a more precise barycentric correction that is calculated with the IDL code ``baryvel.pro'', which is based on the method in \citet{1980A&AS...41....1S} and gives an accuracy of $\sim$1\,\ms. The corrected FEROS spectra are then used in our analysis.

\section{Results}\label{Sec:result}
\subsection{Stellar properties}\label{sec:stellar_properties}
In the literature, the spectral type of GW~Ori ranges from K3 to G0  \citep{1949ApJ...110..424J,1977ApJ...214..747H,1975ascp.book.....H,2004AJ....128.1294C,2010A&A...517A..88W}. We classify GW~Ori by comparing its observed spectra with synthetical spectra. To improve the quality of the observational data, we combine the FEROS spectra\footnote{Before the combination, each spectrum used has been corrected for the Doppler shift by using the value derived in \sect\ref{sect:rv} and normalized to avoid the possible variations in the shape of the spectra.}, which is observed during Jan. 6--11 in 2009, to one spectrum. {\newrev With these data, we do not see any obvious variations in photospheric absorption features among the individual spectra.} For comparison, we used synthetical spectra extracted from \citet{2005A&A...443..735C} with a solar abundance and a surface gravity log~$g$=3.5\footnote{\newrev Using the stellar masses and radius of GW~Ori from \citet{2004AJ....128.1294C}, the surface gravity of GW~Ori is estimated to be log~$g\sim$3.3. Therefore, for simplicity, we only select the synthetical spectra with log~$g$=3.5 from \citet{2005A&A...443..735C}.}.  The comparison spectra were first degraded to match the spectral resolution of FEROS and then rotationally broadened with a rotation velocity $v$~sin~$i_{*}$=43.7\,\kms\ \citep{2010A&A...517A..88W}. 

%GW~Ori is still accreting.
 {\rev The accretion shock can produce excess continuum emissions, which fill the photospheric absorption features and induce a veiling effect on the spectra \citep{1998ApJ...509..802C}.} According to \citet{2004AJ....128.1294C}, the accretion luminosity of GW~Ori is  $\sim$8\% of its stellar luminosity. In addition, the excess continuum emission peaks at the UV band.  Therefore, veiling is expected to be insignificant for GW~Ori in optical bands. However, we still include veilling in the spectral comparison as an independent check. We divide spectra into different wavelength bins with a size of 50\,\AA. Within each bin, we  include a free parameter $r$ to simulate veiling in a  way similar to that in \citet{1989ApJS...70..899H}. The $r$ value is set to be constant within one wavelength bin and can be optimized by minimizing $\chi^2$, as defined by $\chi^2$=$\sum$$(F_{\rm Obs,~\lambda}-\frac{F_{\rm Syn,~\lambda}+r}{1+r})^2$, where $F_{\rm Obs,~\lambda}$ is the flux of the observed spectrum at the wavelength $\lambda$, and $F_{\rm Syn,~\lambda}$ is the flux of the theoretical spectrum at the corresponding wavelength. We find that the GW~Ori spectrum can be  best reproduced by synthetical spectra with effective temperatures ($T_{\rm eff}$) between 5500 and 5750\,K (closer to 5500\,K) and a negligible veiling ($r$$<$0.1), which is consistent with that suggested by \citet{2004AJ....128.1294C}. In Fig.~\ref{Fig:comparison_obs_synthetic}, we show comparisons of the spectra as an example within the wavelength ranges of 4875--4925\AA\ and 5225--5275\AA. Hereafter, we take 5500\,K as the effective temperature of GW~Ori, which corresponds to a spectral type $\sim$G8,  according to the relation between spectral types and $T_{\rm eff}$ that is given in \citet{1995ApJS..101..117K}.

Using our derived spectral type with photometric data in \citet{2004AJ....128.1294C}, we calculate the visual extinction ($A_{\rm V}$) and bolometric luminosity ($L_{\star}$) of the primary of GW~Ori (GW~Ori~A), assuming that the primary dominates the emission of the system. In the calculation, we use the method described in \citet{2009A&A...504..461F} by adopting a total to selective extinction value typical of interstellar medium dust ($R_{\rm V}$=3.1) and  the extinction law from \citet{1989ApJ...345..245C}. The resulting $A_{\rm V}$ and $L_{\star}$ of  GW~Ori~A are 1.5$\pm$0.1\,mag  and 48$\pm$10\,\Lsun, respectively.  We derive the stellar mass ($M_{\star}$) and age of GW~Ori~A using the three sets of publicly available pre-main sequence (PMS) evolutionary tracks from \cite{2000A&A...358..593S} (S00), \cite{2008ApJS..178...89D} (D08), and \cite{2011A&A...533A.109T} (Pisa11). The masses and ages of  GW~Ori~A are 3.7$\pm$0.3\,\Msun\ and 0.9$\pm$0.3\,Myr from S00, 4.0$\pm$0.2\,\Msun\ and 0.4$\pm$0.1\,Myr from D08, and 3.9$\pm$0.2\,\Msun\ and 0.7$\pm$0.2\,Myr from Pisa11. In the following, we use the weighted mean, 3.9$\pm$0.2\,\Msun, as the mass of GW~Ori~A. {\rev With such a high mass, GW~Ori~A would be a B6 main-sequence star \citep{1992A&AS...96..269S}. The late spectral type (G8) of GW~Ori~A suggests that it is at an evolutionary stage earlier than Herbig~Be stars. According to the three PMS evolutionary tracks for a young star with a mass of 4\,\Msun, there is a rapid transition from  G~type to  B~type at an age of $\sim$1\,Myr, which may explain why there are so few known young stars, like GW~Ori~A, with such high masses but with late spectral types.}

\begin{figure}
\centering
\includegraphics[width=1.\columnwidth]{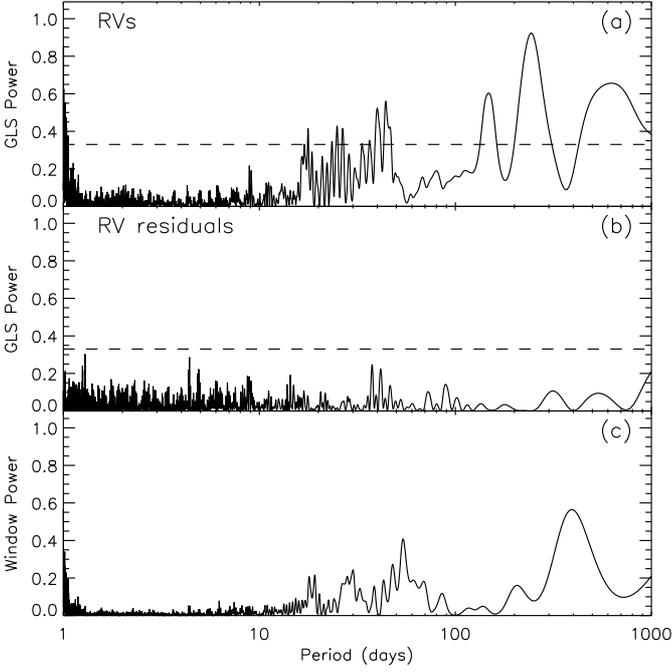}
\caption{(a) Generalized Lomb-Scargle periodogram of the RV measurements of GW~Ori. (b) Generalized Lomb-Scargle periodogram for the RV residuals after subtracting a one-companion fit. (c) Window function. The dashed lines in panels~(a, b) indicate the power level for an FAP of 0.01, as computed by GLS. \label{Fig:allgls}}
\end{figure}

\begin{figure}
\centering
\includegraphics[width=1.\columnwidth]{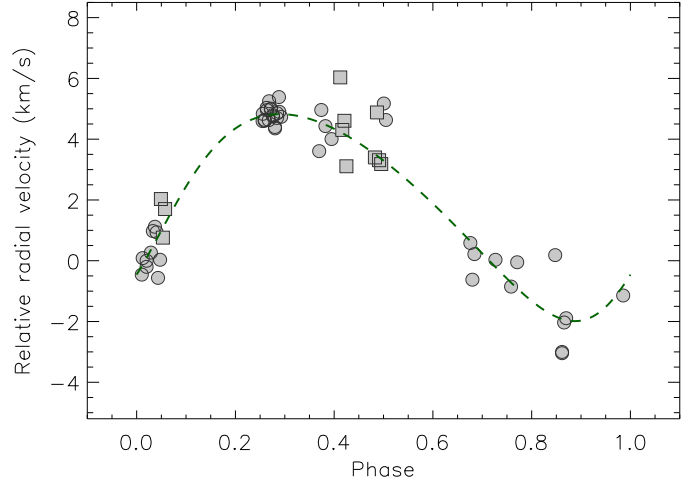}
\caption{Phase-folded relative RV curve (dash line) of GW~Ori. The filled circles show our measurements of the spectra from FEROS, and the filled boxes are for the data from HARPS. \label{Fig:Phase}}
\end{figure}

\begin{table*}
\caption{Relative radial velocity of GW~Ori.\label{Tab:obs_RV}}
\centering
\begin{tabular}{ccccc|ccccc}
\hline\hline
ID  & JD-2450000 &Rel. RV &$\sigma$ (Rel. RV)  & (O-C) &ID  & JD-2450000 &Rel. RV &$\sigma$ (Rel. RV)& (O-C) \\
    &            &(\kms)  &(\kms) &(\kms) &    &            &(\kms)  &(\kms)            &(\kms)\\
\hline
1 & 4412.689 &$+  5.17$&  0.12&$+  1.89$&30& 4844.538 &$+  5.39$&  0.09&
$+  0.56$\\
2 & 4413.727 &$+  4.63$&  0.07&$+  1.40$&31& 4844.680 &$+  4.90$&  0.09&
$+  0.08$\\
3 & 4496.530 &$+  0.18$&  0.08&$+  2.01$&32& 4845.606 &$+  4.73$&  0.07&$ -0.09$
\\
4 & 4777.866 &$+  0.09$&  0.06&$+  0.21$&33& 4874.520 &$+  6.03$&  0.07&
$+  1.56$\\
5 & 4779.777 &$ -0.20$&  0.06&$ -0.30$&34& 4875.532 &$+  4.30$&  0.06&$ -0.13$\\
6$^{a}$ & 4779.788 &$+  0.00$&0.06$^{b}$&$ -0.10$&35& 4876.574 &$+  4.61$&  0.06&
$+  0.21$\\
7 & 4781.869 &$+  0.27$&  0.06&$ -0.09$&36& 4877.543 &$+  3.11$&  0.06&$ -1.25$
\\
8 & 4782.849 &$+  0.98$&  0.06&$+  0.50$&37& 4891.508 &$+  3.40$&  0.06&$ -0.35$
\\
9 & 4783.851 &$+  1.12$&  0.06&$+  0.52$&38& 4892.501 &$+  4.88$&  0.07&
$+  1.18$\\
10 & 4784.690 &$+  0.94$&  0.07&$+  0.23$&39& 4893.519 &$+  3.31$&  0.06&
$ -0.34$\\
11 & 4786.777 &$+  2.03$&  0.06&$+  0.83$&40& 4894.565 &$+  3.18$&  0.06&
$ -0.41$\\
12 & 4787.747 &$+  0.76$&  0.07&$ -0.56$&41& 4950.469 &$+  0.03$&  0.06&
$+  0.24$\\
13 & 4788.813 &$+  1.70$&  0.07&$+  0.24$&42& 5105.810 &$+  3.61$&  0.25&
$ -0.95$\\
14 & 4836.533 &$+  4.59$&  0.06&$ -0.16$&43& 5106.849 &$+  4.96$&  0.18&
$+  0.43$\\
15 & 4836.629 &$+  4.83$&  0.10&$+  0.07$&44& 5108.867 &$+  4.43$&  0.12&
$ -0.04$\\
16 & 4837.538 &$+  4.61$&  0.06&$ -0.17$&45& 5111.876 &$+  4.00$&  0.09&$ -0.38$
\\
17 & 4837.599 &$+  4.63$&  0.07&$ -0.14$&46& 5179.775 &$+  0.58$&  0.06&$ -0.06$
\\
18 & 4838.538 &$+  5.03$&  0.07&$+  0.24$&47& 5180.777 &$ -0.62$&  0.06&$ -1.20$
\\
19 & 4838.689 &$+  4.96$&  0.11&$+  0.17$&48& 5181.762 &$+  0.22$&  0.06&
$ -0.29$\\
20 & 4839.567 &$+  4.63$&  0.08&$ -0.17$&49& 5199.705 &$ -0.85$&  0.06&$ -0.13$
\\
21 & 4839.720 &$+  5.25$&  0.08&$+  0.45$&50& 5202.702 &$ -0.05$&  0.06&
$+  0.86$\\
22 & 4840.560 &$+  4.98$&  0.08&$+  0.17$&51& 5224.650 &$ -3.04$&  0.14&$ -1.13$
\\
23 & 4840.687 &$+  5.01$&  0.07&$+  0.20$&52& 5224.663 &$ -3.00$&  0.07&$ -1.08$
\\
24 & 4841.542 &$+  4.79$&  0.08&$ -0.02$&53& 5225.664 &$ -2.03$&  0.13&$ -0.10$
\\
25 & 4841.632 &$+  4.75$&  0.07&$ -0.06$&54& 5226.644 &$ -1.89$&  0.09&$+  0.06$
\\
26 & 4842.572 &$+  4.40$&  0.10&$ -0.42$&55& 5254.566 &$ -1.14$&  0.06&$ -0.32$
\\
27 & 4842.633 &$+  4.36$&  0.06&$ -0.46$&56& 5260.595 &$ -0.46$&  0.06&$ -0.28$
\\
28 & 4843.538 &$+  4.70$&  0.09&$ -0.12$&57& 5268.535 &$ -0.56$&  0.10&$ -1.35$
\\
29 & 4843.619 &$+  4.85$&  0.08&$+  0.03$&58& 5269.541 &$+  0.03$&  0.10&
$ -0.88$\\
\hline
\end{tabular}
\tablefoot{\rev Column~3: RVs relative to the ID~6 spectrum. Column~5: RV residuals from the orbital solution. a: Template used for deriving the relative RVs of other spectra. b: Typical error for other spectra.}
\end{table*}

\subsection{Radial velocity, Keplerian orbital solution, and rotation}\label{sect:rv}

We derive the radial velocity (RV) of each spectrum of GW~Ori using the cross-correlation method by taking one observed spectrum (ID~6 in Table~\ref{Tab:obs_log}) as the template. The relative RVs are listed in Table~\ref{Tab:obs_RV}. In Fig.~\ref{Fig:allgls}(a), we show the  generalized Lomb-Scargle (GLS) periodogram of the RVs that are calculated using the method in \citet{2009A&A...496..577Z}. The GLS periodogram shows three strong peaks at $\sim$120, 240, and 600\,days, respectively. The strongest peak is at $\sim$240\,days with a false-alarm probability (FAP) of 1.7$\times$10$^{-28}$ estimated by GLS.

\begin{table}
\caption{Orbital elements for GW~Ori.\label{Tab:orbit_par}}
\centering
\renewcommand{\tabcolsep}{0.01cm}
\begin{tabular}{ccc}
\hline\hline
Parameters &This work & In \citet{1991AJ....101.2184M}\\
$P$ (days)&$     241.6\pm       1.5$ &241.9$\pm$1.0\\
$T_{\rm p}$&$    4774.9\pm       8.0$ & \\
(JD-2450000)\\
e&$ 0.18\pm 0.06$ &0.04$\pm$0.06\\
$\omega$ (deg)&$     236.6\pm      13.4$ &71$\pm$60\\
$K$ (\kms)&$ 3.41\pm 0.17$ &4.7$\pm$0.3\\
$f(m)$ (\Msun)&$(9.39\pm1.16)\times10^{-4}$ &2.6$\pm0.5\times10^{-3}$\\
$\sigma$ (O$-$C) (\kms)&0.7 &1.3\\
\hline
With $m_{1}=3.9\pm0.2$\Msun\\
$m_{2}$~sin~$i$ (\Msun)&$ 0.25\pm 0.01$ \\
$a_{2}$ (AU)&$ 1.20\pm 0.02$ \\
\hline
\end{tabular}
\end{table}
 % treated, RvB

We achieve a one-companion Keplerian orbital solution for GW~Ori by using the fitting procedure in \citet{2009ApJS..182..205W} and estimate the  uncertainties of parameters by using the bootstrapping routines in \citet{2012ApJ...761...46W}. A  phase-folded RV curve of GW~Ori is shown in Fig.~\ref{Fig:Phase}. The best-fit orbital parameters  are summarized in Table~\ref{Tab:orbit_par}, which includes the orbital period $P$, a Julian date of periastron passage  $T_{\rm P}$, the eccentricity $e$, the periastron angle $\omega$, the semi-amplitude of the RV curve $K$, the mass function $f(m)$, and the rms residual velocities from the orbital solution. In Table~\ref{Tab:orbit_par}, we also present the orbital parameters from \citet{1991AJ....101.2184M} as a comparison. We note our rms residual velocity is about half of that in \citet{1991AJ....101.2184M}. Both works agree with each other on the orbital periods. Our eccentricity is marginally larger than the previous one. The largest differences between both works arise mostly in $\omega$, $K$, and $f(m)$. According to the uncertainties of $\omega$, our data may provide a better constraint on $\omega$ than \citet{1991AJ....101.2184M}. Taking 3.9$\pm$0.2\,\Msun\ as the mass of GW~Ori~A, we obtain a minimum companion mass ($m_{2}$~sin~$i$) $\sim$0.25\,\Msun, and an orbital semi-major axis ($a$) of 1.20\,AU. The $a$ value is consistent with the one in \citet{1991AJ....101.2184M}.

\begin{figure}
\centering
\includegraphics[width=1.\columnwidth]{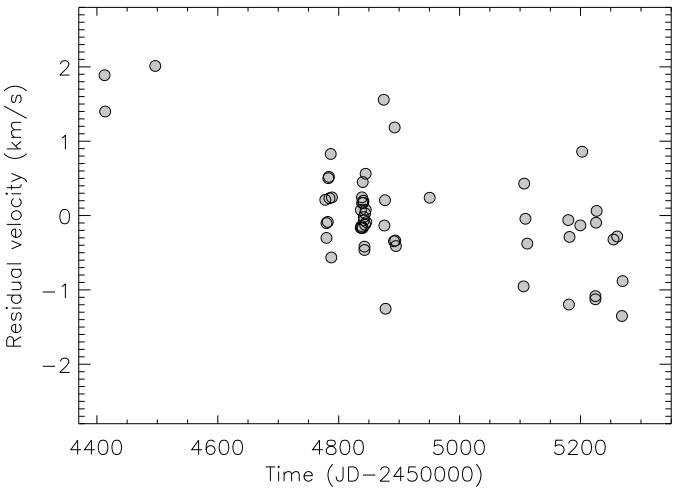}
\caption{The residual velocities of GW~Ori plotted as a function of observational dates. \label{Fig:residual_longperiod}}
\end{figure}

\begin{figure}
\centering
\includegraphics[width=1.\columnwidth]{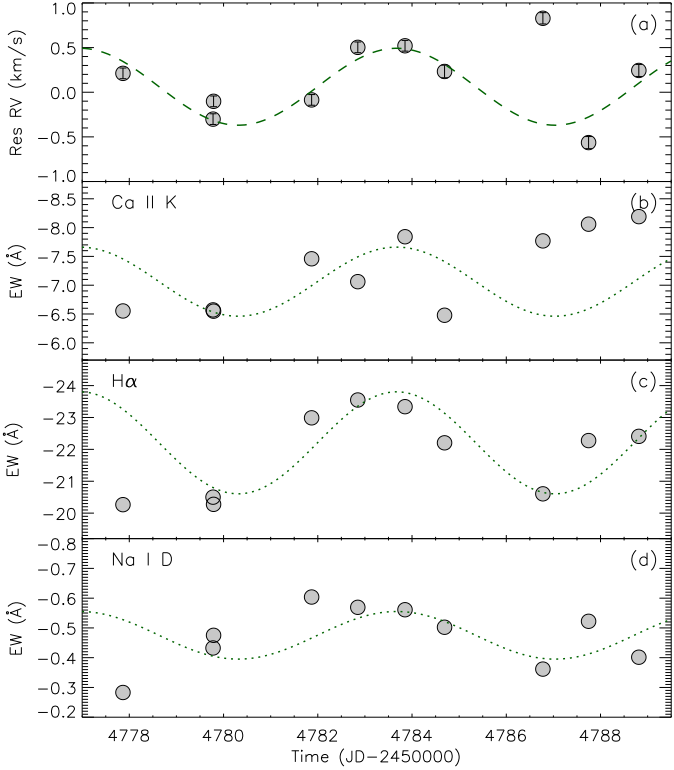}
\caption{(a) The residual velocities of GW~Ori. The error bars of individual measurements are shown. The dashed line is the fit to the residual RVs. The dotted lines in panels~(b, c, d) are sine functions with  the fitted phase and period from panel~(a). We manually shift the sine functions and adjust the amplitudes to match the observed EWs of each emission line. \label{Fig:residual1}}
\end{figure}

\begin{figure}
\centering
\includegraphics[width=1.\columnwidth]{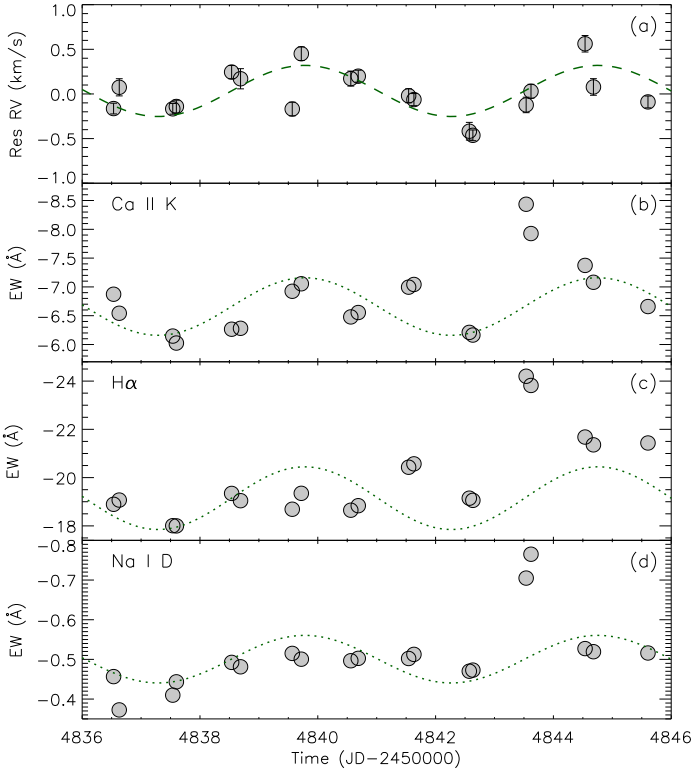}
\caption{Same as in Fig.~\ref{Fig:residual1} but for different observational dates.  \label{Fig:residual2}}
\end{figure}

\begin{figure}
\centering
\includegraphics[width=1.\columnwidth]{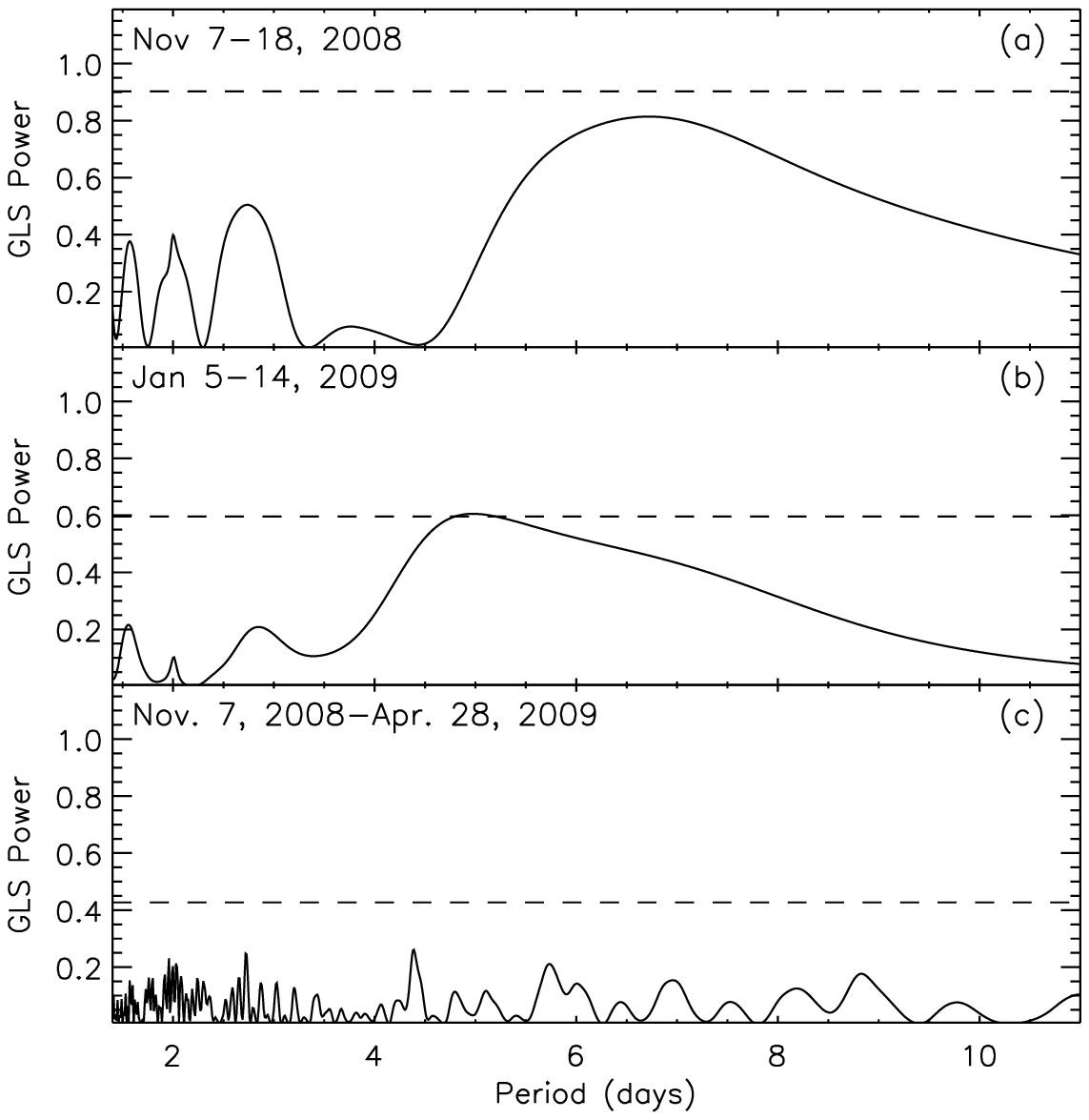}
\caption{ (a) Generalized Lomb-Scargle periodogram of the RV residuals of GW~Ori during Nov. 7--11, 2008. (b) The same periodogram for the  RV residuals but for data obtained during Jan. 5--14, 2009. (c) Same as in panels~(a, b) but for data taken between Nov. 7, 2008 and Apr. 28, 2009. The dashed line in each panel indicates the power level for an FAP of 0.01, as computed by GLS.  \label{Fig:rogls}}
\end{figure}

Table~\ref{Tab:obs_RV} lists the individual RV residuals from the orbital solution. The rms residual velocity is $\sim$0.7\,\kms, which considerably higher than internal measurement errors ($\lesssim$0.1\,\kms). We show the GLS periodogram of the RV residuals in Fig.~\ref{Fig:allgls}(b). We note that the two peaks (120 and 600\,days) shown in Fig.~\ref{Fig:allgls}(a) disappear, and no significant periods with FAP$<$0.01 are present in the periodogram. In Fig.~\ref{Fig:residual_longperiod}, we show the RV residuals with respect to observational dates and note a systematic shift in the RV residuals with the time. \citet{1991AJ....101.2184M} also found a systematic shift in their RV residuals and explained it as being caused by a second companion (GW~Ori~C). The GW~Ori~C has been confirmed with the infrared interferometric technique by \citet{2011A&A...529L...1B} and  is located at a project separation $\sim$8\,AU from GW~Ori~A, indicating an orbital period $\sim$3600\,days. Our observations only have a time span of $\sim$850\,days and, thus, cannot provide any useful constrains on the orbital parameters of GW~Ori~C.

\begin{figure*}
\centering
\includegraphics[width=1.9\columnwidth]{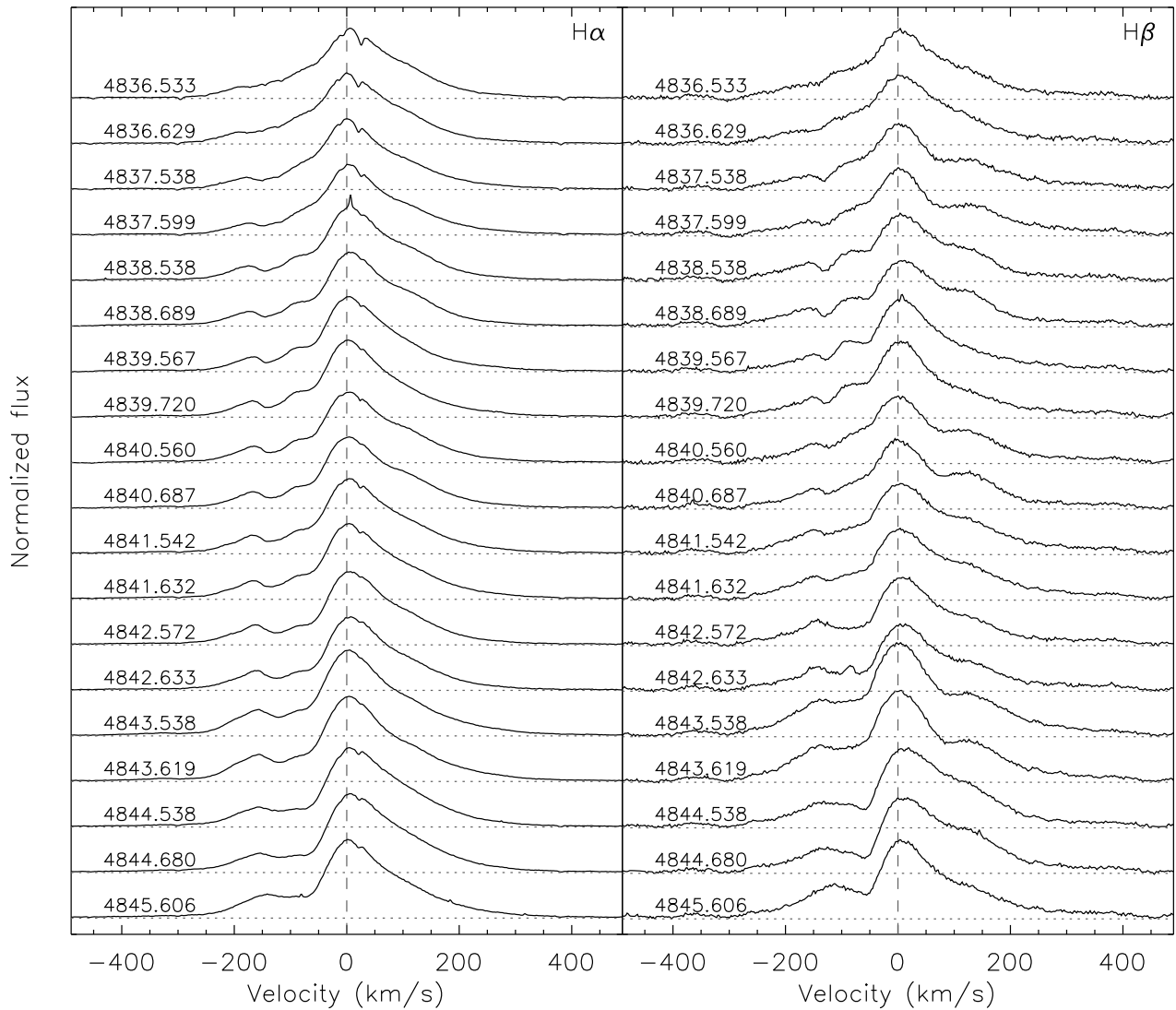}
\caption{Residual line profiles of H$\alpha$ and H$\beta$ observed during Jan 5--14. The profiles have been shifted for clarity. The vertical dashed lines mark the spectral line center at the stellar rest frame, and the horizontal dotted lines show the continuum level. The numbers on the left are the Julian dates of observation minus 2450000.}\label{Fig:epoch_allHalpa}
\end{figure*}

In our datasets, there are two time intervals, Nov. 7--18, 2008 and Jan. 5--14, 2009, with very dense time coverage. In Figs.~\ref{Fig:residual1} and \ref{Fig:residual2}, we show the RV residuals during the two periods and note periodic variations in the RV residuals. We fit the data points ($v_{\rm res}$)\footnote{In each fit, we have excluded one data point: ID\,11 in Fig.~\ref{Fig:residual1} and ID\,20 in Fig.~\ref{Fig:residual2}. The two  data points seem to deviate very much from the global trends of the $v_{\rm res}$ variations. The two data points are also excluded when we calculate the GLS periodograms in Fig.~\ref{Fig:rogls}~(a, b)} by using the form  $v_{\rm res}=v_{\rm max}\times sin(\frac{2\pi t}{\tau}+\theta)+v_{0}$,  where $v_{0}$ is the systematic shift caused by the second companion, and  $\tau$ is the period of the $v_{\rm res}$ variation. The best-fit results with $\tau$ $\sim$6.7 and 5.0\,days are also shown in Fig.~\ref{Fig:residual1} and~\ref{Fig:residual2}, respectively. {\rev We calculate the GLS periodograms of the RV residuals for the data taken during the two time intervals and show them in Fig.~\ref{Fig:rogls}~(a, b). The peaks of $\sim$6.7 and 5.0 can be noted in the two periodograms with FAP$\sim$0.07 and 0.01, respectively. In Fig.~\ref{Fig:rogls}~(c), we show the GLS periodogram for the data taken between Nov. 7, 2008 and Apr. 28, 2009. We only use the data  during that time because the systematic shifts caused by the second companion are small (see Fig.~\ref{Fig:residual_longperiod}).  In Fig.~\ref{Fig:rogls}~(c), no strong peaks can be noted, suggesting the periodic variations shown in Fig.~\ref{Fig:residual1} and~\ref{Fig:residual2} are not stable. We also note that the best-fit functions for the two sets of data are inconsistent in both amplitudes and phases. Thus, the period of 5.0-6.7~days cannot be due to  a new companion. One promising explanation could be rotational modulation \citep[see e.g.][]{2001A&A...379..279Q,2008ApJ...687L.103P,2011A&A...530A..85M}.  The disappearance and appearance of the spots, as well as the variations of filling factors of spots on the stellar surface, can lead to the change in both amplitudes and phases in the periodic variations of the RV residuals due to the rotational modulation.}

During these two time intervals, we find that the equivalent widths (EW) of accretion-related emission lines exhibit  similar periodic variations to the RV residuals, although less distinct. Figures~\ref{Fig:residual1} and~\ref{Fig:residual2} show the EWs of three accretion-related emission lines, Ca\,II~K line, H$\alpha$, and Na\,I\,D line, with respect to the observational dates as examples (see \sect~\ref{Sec:emission_lines} for a detail description of emission lines in the spectra of GW~Ori). The periodic variation of the EWs of accretion-related emission lines can be also explained as a rotational modulation since the filling factor of the accretion-shock region on the stellar surface is small, and the accretion streams are usually not azimuthally symmetrically distributed with respect to the stellar rotational axis \citep{1995ApJ...449..341J,2004ApJ...610..920R,2007prpl.conf..479B}. In addition to the rotational modulation, the EW variation of accretion-related emission lines can also be caused by accretion variation, such as non-steady accretion on the timescale of hours, the global instabilities of the magnetospheric structure on the timescale of months, or the pulsed accretion due to the orbital motion for binaries \citep{1996A&A...314..835G,2003A&A...409..169B,1996ApJ...467L..77A}. These factors can contaminate the periodic behaviors of EWs of the emission lines in Figs.~\ref{Fig:residual1} and~\ref{Fig:residual2}.

{\rev Taking the stellar radius 7.6\,$R_{\odot}$, the break-up velocity of GW~Ori~A is estimated to be 442\,\kms, and its lower-limit rotational period is 0.9~days. According to our derived rotational period (5.0--6.7\,days), GW~Ori~A is far away from the limit.} Given $v$~sin~$i_*$=43.7\,\kms, we calculate the inclination ($i_*$) of the stellar rotation axis of GW~Ori~A, which is around 35--50$^{\circ}$. \citet{1990AJ.....99..946B} derived a shorter rotation period ($\sim$3.3\,days) for GW~Ori~A mainly based on the $U$-band photometry. It is unknown which period is more accurate, since both our data and theirs are not sampled very well over the time. If the rotational period of GW~Ori~A was the shorter one, the inclination would be around 22$^{\circ}$. Here, we consider that the inclination of the rotational axis is between 22--50$^{\circ}$, {\subrev although current data cannot give any constraints on the inclination of the  orbital axis of GW~Ori~A/B.} If the orbital axis of GW~Ori~A/B is aligned with the rotational axis of GW~Ori~A, the mass of GW~Ori~B is  0.3--0.7\,\Msun. {\subrev However, an intermediate inclination of  the orbital axis of GW~Ori~A/B contradicts the observed eclipses by \citet{1998AstL...24..528S}. The eclipses were detected during 1987--1992 and then disappeared \citep{1998AstL...24..528S}. Therefore, more data are required to understand the nature of the eclipses and, furthermore, to give a constraint on the inclination of the  the orbital axis of GW~Ori~A/B.}

\begin{figure}
\centering
\includegraphics[width=1\columnwidth]{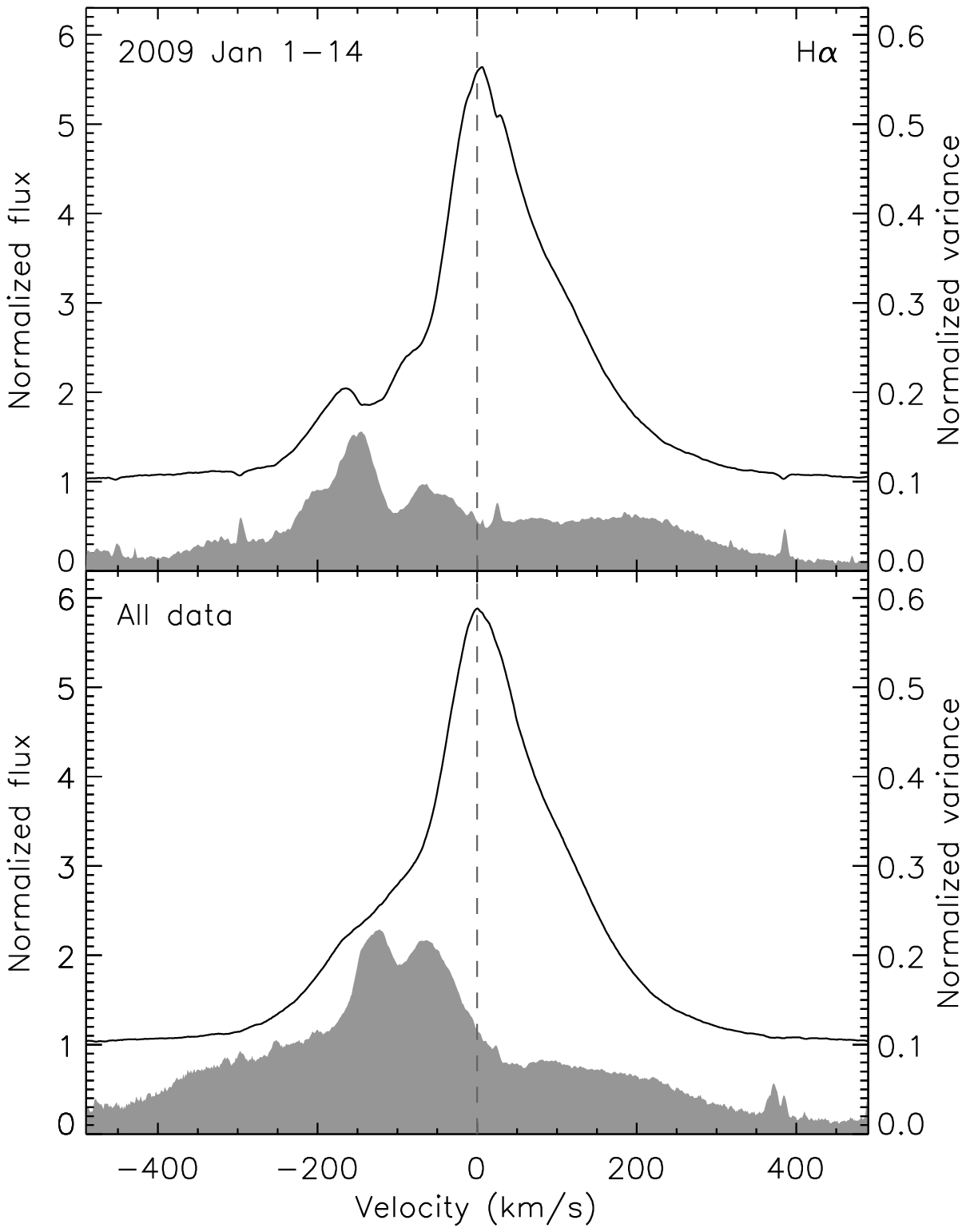}
\caption{Average H$\alpha$ line profiles (solid line) and normalized variance profiles (gray shaded areas) calculated with residual spectra. The vertical dashed lines mark the spectral line center at the stellar rest frame.}\label{Fig:allHalpa_variance}
\end{figure}

\begin{figure}
\centering
\includegraphics[width=1\columnwidth]{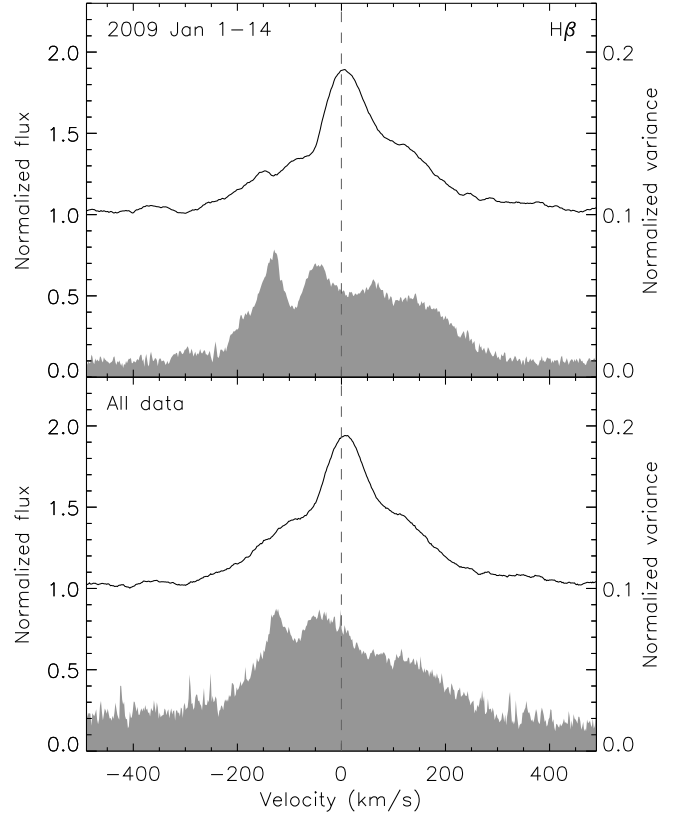}
\caption{Similar to Fig.~\ref{Fig:allHalpa_variance} but for H$\beta$ line.}\label{Fig:allHbeta_variance}
\end{figure}

\subsection{Emission lines}\label{Sec:emission_lines}

\subsubsection{H$\alpha$ and H$\beta$ emission lines}\label{Sec:Halpha_Hbeta_lines}

\vspace{0.3cm} 
\noindent 
\noindent \textbf{(a) Equivalent widths and line profiles}\\ 
\vspace{-0.3cm}

\begin{figure}
\centering
\includegraphics[width=1\columnwidth]{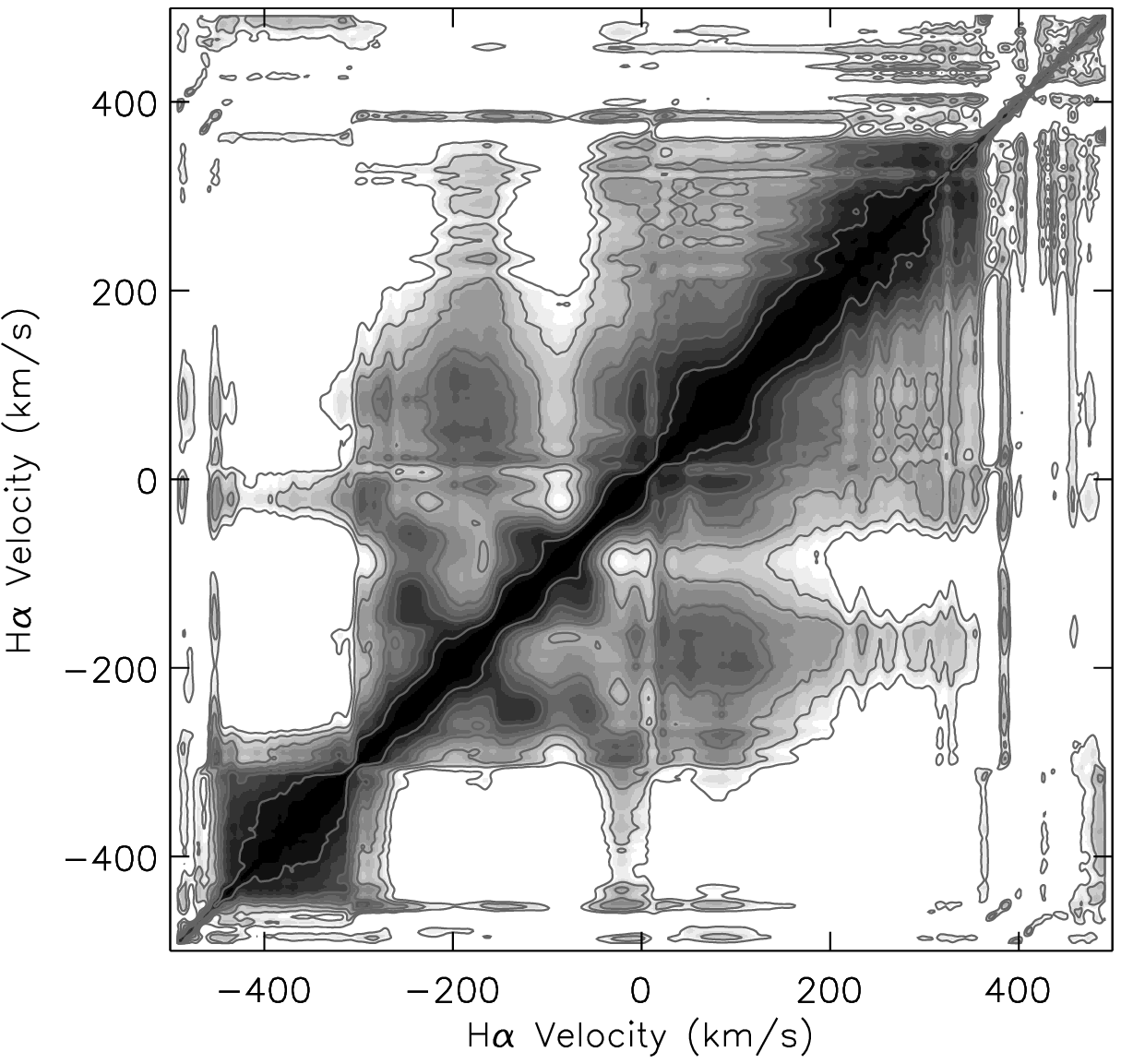}
\includegraphics[width=1\columnwidth]{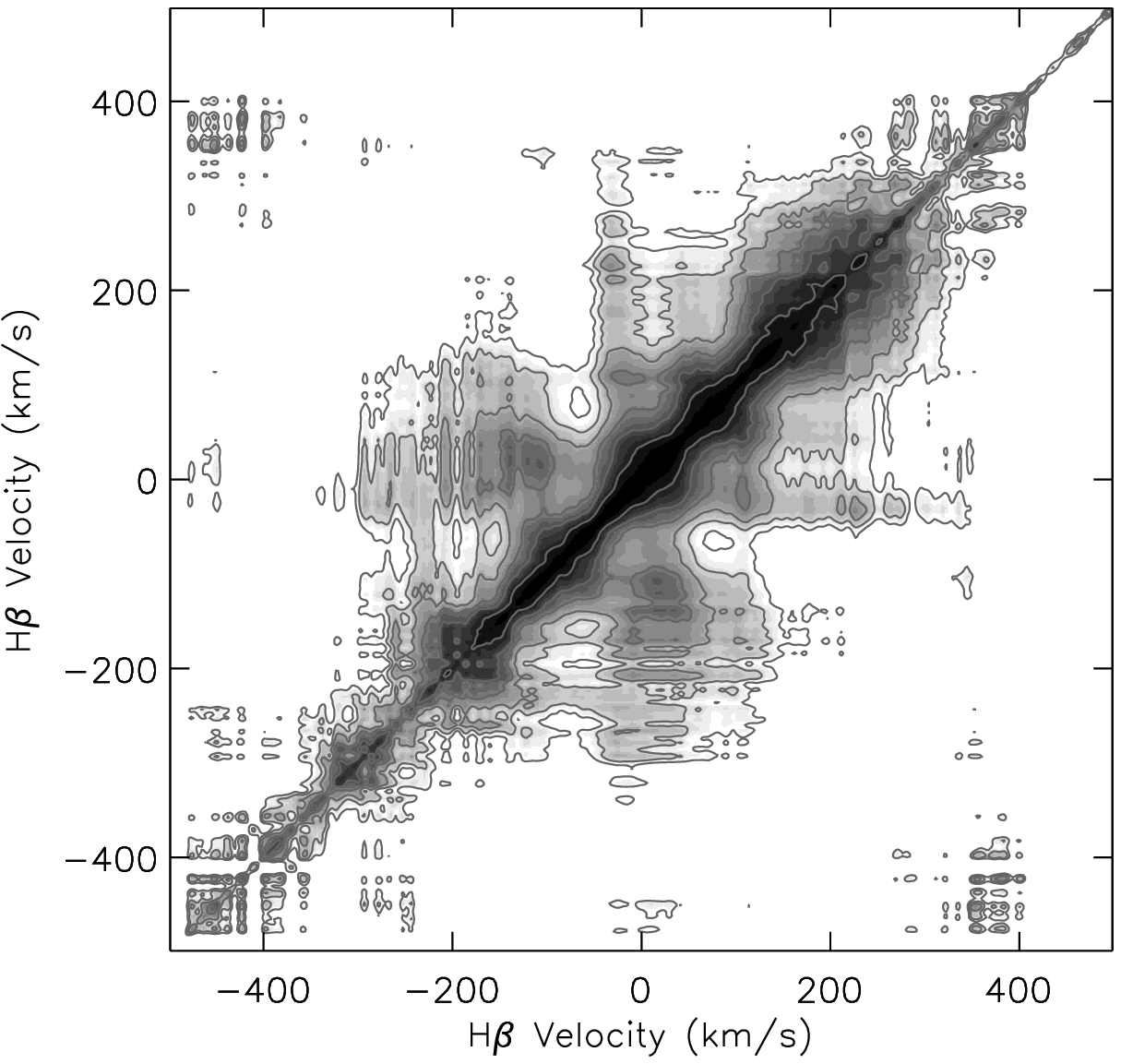}
\caption{Correlation matrices of H$\alpha$ and H$\beta$ lines for GW~Ori. The lowest contours in each panel corresponds to 99.9\% confidence level. \label{Fig:Halpa_Hbeta_cc}}
\end{figure}

\begin{figure}
\centering
\includegraphics[width=1\columnwidth]{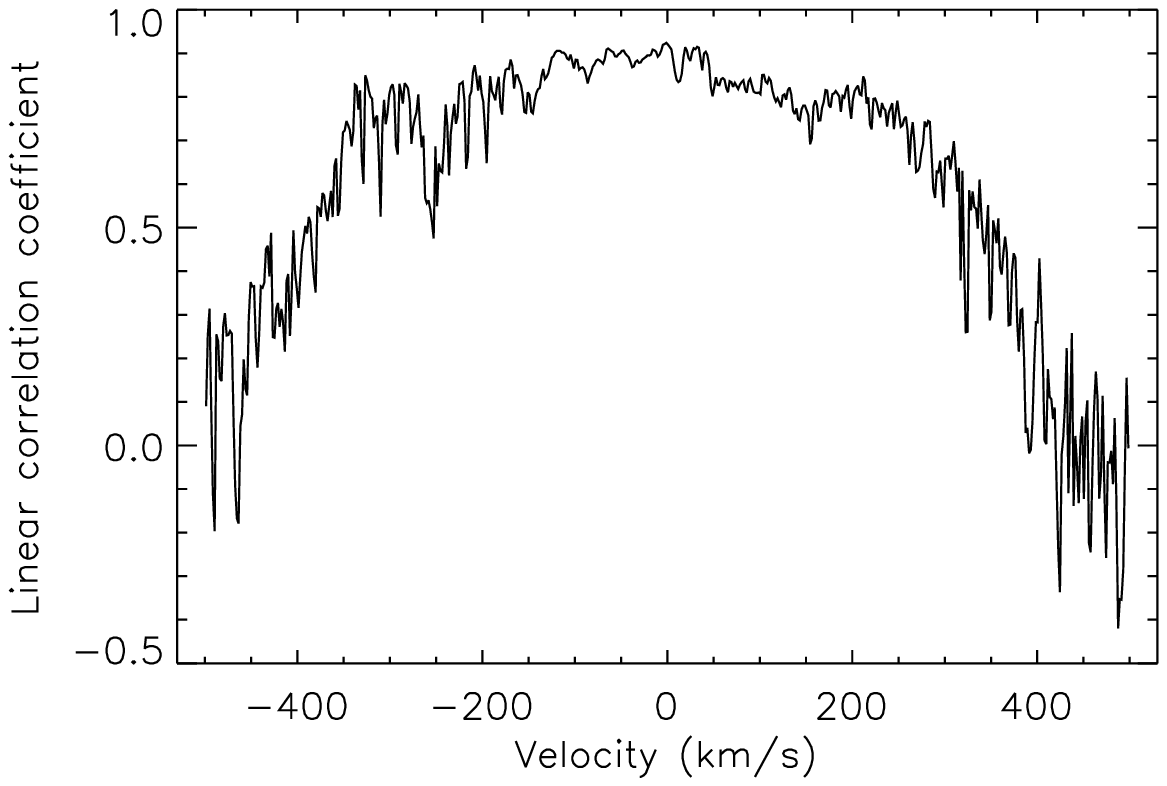}
\caption{The linear correlation coefficient between H$\alpha$ and H$\beta$ line profiles in each velocity bin.}\label{Fig:Halpa_Hbeta_line}
\end{figure}

The H$\alpha$ and  H$\beta$ lines are prominent in our spectra of GW~Ori due to their strong and broad line profiles. In Fig~\ref{Fig:epoch_allHalpa}, we show the example residual profiles of H$\alpha$  and  H$\beta$  lines observed during Jan 5--14, 2009. The residual line profiles are obtained by normalizing the observed spectra and subtracting the photospheric absorption features from the normalized synthetical spectrum with $T_{\rm eff}$=5500\,K. As shown in Fig~\ref{Fig:epoch_allHalpa}, the H$\alpha$ and  H$\beta$ lines show very similar profiles. In both lines, the most notable feature is a blue-shifted absorption, which changes in both strength and central velocity with time. In Table~\ref{Tab:EW_line}, we list the EWs of H$\alpha$  and  H$\beta$ calculated with the residual line profiles.

\vspace{0.3cm} 
\noindent 
\noindent \textbf{(b) Line variance}\\ 
\vspace{-0.3cm} 
 
In Figs.~\ref{Fig:allHalpa_variance} and \ref{Fig:allHbeta_variance}, we show the average residual profiles of H$\alpha$ and H$\beta$ lines with the normalized variance profiles, as calculated with the data observed during 2009 Jan 1--14, and all the observations. The variance profiles, as defined in \citet{1995AJ....109.2800J}, are  measurements of variability in each velocity bin within the lines. The variance profiles of  H$\alpha$ and H$\beta$ show blue-shifted peaks, corresponding to the variable absorption features (see Fig.~\ref{Fig:epoch_allHalpa}), and are featureless on the red side.

\vspace{0.3cm} 
\noindent 
\noindent \textbf{(c) Correlation matrices}\\ 
\vspace{-0.3cm} 
 
We calculate autocorrelation matrices for the H$\alpha$ and H$\beta$ lines and use them to investigate how the variations of line profiles are correlated across the lines \citep[see][]{1995AJ....109.2800J}. The resulting matrices are  illustrated in Fig.~\ref{Fig:Halpa_Hbeta_cc}. For H$\alpha$, the autocorrelation matrix shows a clear correlation between the blue ($-$300--0\,\kms) and the red (0--200\,\kms) sides of profiles. The autocorrelation matrix of H$\beta$ exhibits a similar pattern to that of H$\alpha$ but with less significance. In Fig.~\ref{Fig:Halpa_Hbeta_line}, we display the linear correlation coefficient between the H$\alpha$ and H$\beta$ profiles within each velocity bin. The H$\alpha$  and H$\beta$ variations are correlated well across the whole line profiles ($-$400--400\,\kms).

\vspace{0.3cm} 
\noindent 
\noindent \textbf{(d) Decomposition of the H$\alpha$ and H$\beta$ line profiles}\\ 
\vspace{-0.3cm}

\begin{figure}
\centering
\includegraphics[width=1.\columnwidth]{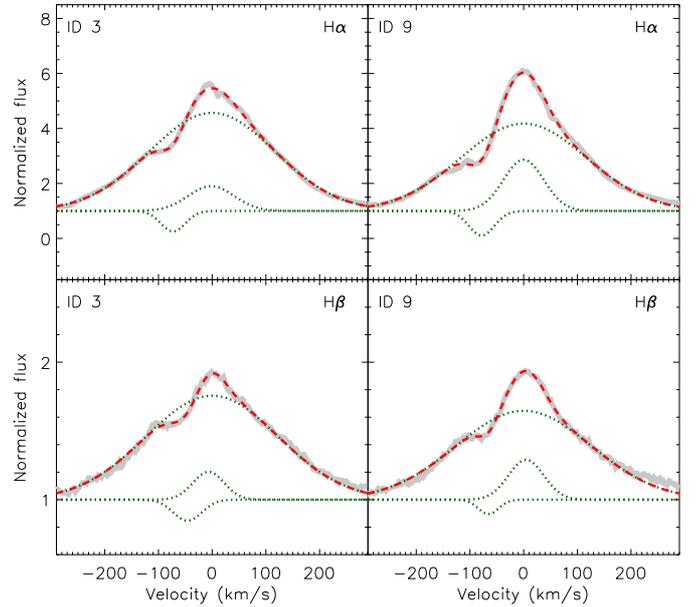}
\caption{Examples of decomposition of H$\alpha$ and H$\beta$ line profiles using three Gaussian components. In each panel, the thick gray line is the observed spectrum; the dotted lines display individual components, and the dashed line shows the final fit. The zero velocity is that of the GW Ori rest frame.}\label{Fig:Halpa_decomposition}
\end{figure}

We decompose the H$\alpha$ and H$\beta$ line profiles of GW~Ori using multi-Gaussian functions, as done in \citet{2012A&A...544A..93S}. We find that all the H$\alpha$ and H$\beta$ line profiles can be fittted well with two emission components, of which one  is strong and broad and the other is narrow and weak, and one blue-shifted absorption component. Figure~\ref{Fig:Halpa_decomposition} shows the examples of fits to the H$\alpha$ and H$\beta$ line profiles. For H$\alpha$ line, the two emission components always peak at $\sim$0\,\kms\ with respect to the rest frame of GW~Ori and show mean EWs of $-$21.6$\pm$2.9 and $-$3.5$\pm$0.6\,\AA, respectively, for the broad and narrow components. The full widths of H$\alpha$ at 10\% ($FW_{\rm H\alpha, 10\%}$) of the peak intensity are 509$\pm$13 and 156$\pm$16\,\kms\ for the two components. There are no correlations between the two emission components in both the EWs and line widths, indicating they may originate from different physical processes. According to the criteria for distinguishing accretors and non-accretors using $FW_{\rm H\alpha, 10\%}$ and EW of H$\alpha$ in \citet{2009A&A...504..461F,2013ApJS..207....5F}, the narrow H$\alpha$ emission component could be mainly related to the chromospheric activity, while the broad one should be due to the accretion activity.

{\rev The two emission components from  H$\beta$ decomposition are similar to those from  H$\alpha$ and peak around  $\sim$0\,\kms\ with respect to the rest frame of GW~Ori. For the broad component, the mean full width of H$\beta$ at 10\% ($FW_{\rm H\beta, 10\%}$) of the peak intensity is 541$\pm$25\,\kms, and the mean EW is  $-$3.4$\pm$0.6\,\AA. The two values are 128$\pm$17\,\kms\ and $-$0.4$\pm$0.1\,\AA, respectively, for the narrow emission component. We note that  there are correlations between the  H$\alpha$ and  H$\beta$ lines in both  emission components. To test the significance of the correlation between them, we apply a Kendall $\tau$ test. If two datasets are fully correlated, the test returns a value of $\tau=1$. If they are anti-correlated, we get $\tau=-1$, and if they are independent, $\tau$ has a value of 0. {\subrev The Kendall $\tau$ test also returns a probability $p$, which is smaller when the correlation is more significant.} For the broad components, the Kendall $\tau$ test yields $\tau$=0.45 and $p$=5$\times$10$^{-7}$ for $FW_{\rm H\alpha, 10\%}$ and  $FW_{\rm H\beta, 10\%}$, and $\tau$=0.62 and $p$=0 for the EWs of both lines. The correlations for the narrow components of the H$\alpha$ and  H$\beta$ lines are less significant with  $\tau$=0.30 and $p$=1$\times$10$^{-3}$ from the Kendall $\tau$ test for $FW_{\rm H\alpha, 10\%}$ and  $FW_{\rm H\beta, 10\%}$ and $\tau$=0.38 and $p$=3$\times$10$^{-5}$ for the EWs of both lines.}

\begin{figure}
\centering
\includegraphics[width=1.\columnwidth]{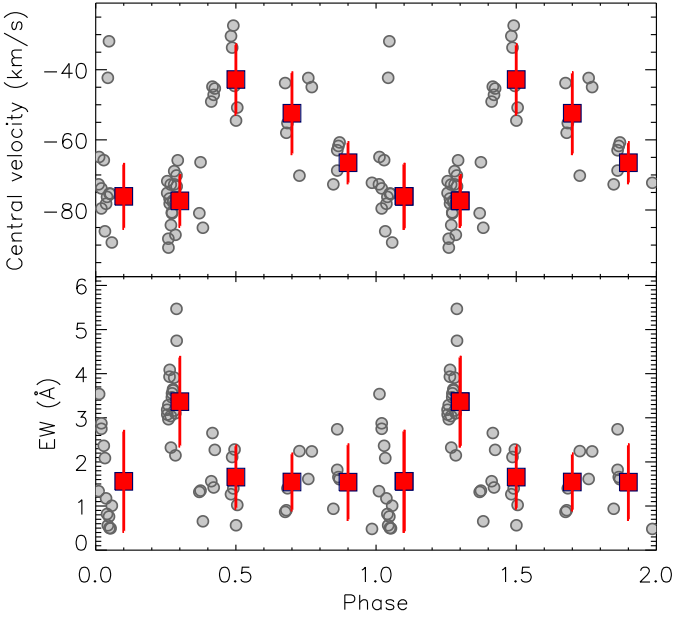}
\caption{(a) Central velocities of the disk wind components from the decomposition of H$\alpha$ line profiles, as plotted over the orbital phases of GW~Ori~B. The velocity is in the rest frame of GW~Ori. We divide the orbital phases into different bins and show the mean value of central velocities in each bin with a filled square and its standard deviation with an error bar. (b) EWs of the disk wind components, as shown in panel~(a), plotted over the orbital phases of GW~Ori~B. We divide the orbital phases into different bins and show the mean value of EWs in each bin with a filled square and its standard deviation with an error bar.}\label{Fig:Halpa_wind}
\end{figure}

{\rev The blue-shifted absorption components in the H$\alpha$ and H$\beta$ line profiles can be related to the wind activity of GW~Ori. The absorption components are variable in both the central velocities and the line strengths for both  H$\alpha$ and H$\beta$ lines, which induces the peaks on the variance at the blue side of their line profile (see Fig.~\ref{Fig:allHalpa_variance} and ~\ref{Fig:allHbeta_variance}). We note that the absorption components in the two lines are strongly correlated in both the central velocities and the EWs.  The Kendall $\tau$ test yields $\tau$=0.53 and $p$=0 for their central velocities and $\tau$=0.57 and $p$=0 for their EWs.} In Fig.~\ref{Fig:Halpa_wind}, we show the central velocities and EWs for the absorption components in H$\alpha$ line profiles with respect to the orbital phases of GW~Ori~B. We note the variations of the two values seem to be periodic with a period comparable to orbital period of GW~Ori~B. This periodic variation of the central velocity and the line strengths is more notable from the mean values within individual phase bins, which are also shown in Fig.~\ref{Fig:Halpa_wind}. The median central velocity of the wind component is around $-$70\,\kms, which is much smaller than the escape velocity ($\sim$440\,\kms) at the surface of GW~Ori~A and comparable to the escape velocity from the disk at a distance of  1.4\,AU. Thus, the blue-shifted absorption component should be attributed to disk winds launched near the orbit of GW~Ori~B and could be affected by the orbital motion of GW~Ori~B. Furthermore, we also note that the EWs of blue-shifted absorption components in Fig.~\ref{Fig:Halpa_wind} are usually larger when they are bluer, which can be due to that wind speeds and EWs are usually correlated in powerful winds. {\rev The blue-shifted absorption components in H$\beta$ line profiles show similar variations with the orbital phases of GW~Ori~B to those of H$\alpha$ lines.}

\begin{figure*}
\centering
\includegraphics[width=1.9\columnwidth]{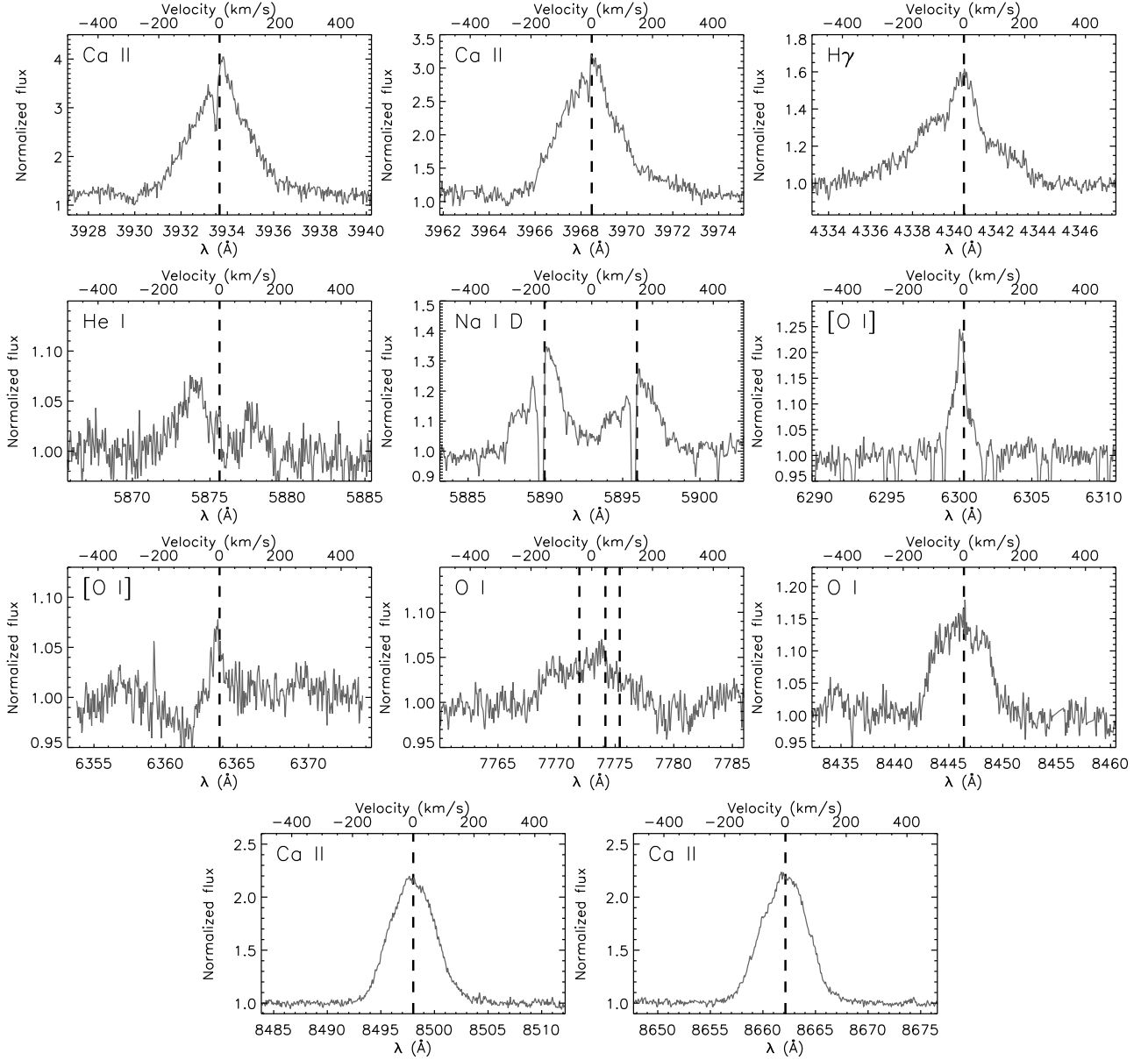}
\caption{The example profiles of the emission lines detected in the optical spectrum (ID\,1 in Table~\ref{Tab:obs_log}) of GW~Ori. The vertical dashed lines mark  the spectral line center at the stellar rest frame. For the O\,I triplet at 7773\,\AA, the central wavelength for each line is marked. The central absorption components in Ca\,II at 3933 and 3968\,\AA\ and in Na\,I at 5890/5896\,\AA\ are due to the absorption from interstellar medium.}\label{Fig:allline}
\end{figure*}

\subsubsection{Other emission lines}
Besides H$\alpha$ and H$\beta$, we also detect other emission lines in the spectra of GW~Ori, including the H$\gamma$ Balmer line, Ca\,II lines at 3933, 3968, 8498, and 8662\,\AA\footnote{The Ca II line at 8542\,\AA\ falls at the edge of the FEROS gap and is thus excluded it here.}, He\,I\,$\lambda$5876, O\,I lines at 6300, 6363, 7773, and 8446\,\AA, and Na\,I D lines at 5890/5896\,\AA. In Fig.~\ref{Fig:allline} we show the residual profiles of these emission lines obtained on  Nov. 8, 2007 as an example. All lines except the oxygen doublet at 6300 and 6363\,\AA\ exhibit broad profiles. In Table~\ref{Tab:EW_line}, we list the EWs of these  lines calculated  {\newrev from their residual line profiles (see the definition of residual line profiles in \sect~\ref{Sec:Halpha_Hbeta_lines}(a)}). 

The H$\gamma$ line in GW~Ori is much weaker than H$\alpha$ and H$\beta$ and appears in emission only after the subtraction of the photospheric absorption. However, the residual profiles of the H$\gamma$ lines are very similar to those of the H$\alpha$ and H$\beta$ lines. In Fig.~\ref{Fig:allline}, the He\,I\,$\lambda$5876 line clearly shows double peaks centered at  $\sim-$100\,\kms\ and  $\sim$100\,\kms. This kind of He\,I\,$\lambda$5876 line profile is quite atypical among T~Tauri stars \citep[see e.g.,][]{1998AJ....116..455M,2001ApJ...551.1037B}. In Fig.~\ref{Fig:allline}, the Ca\,II lines at 3933, 3968, 8498, and 8662\,\AA\ and the Na\,I D lines at 5890/5896\,\AA\ show  broad line profiles with blueward asymmetry. All these above described lines have been widely used as tracers of accretion in the literature \citep{2005ApJ...626..498M,2008ApJ...681..594H,2009A&A...504..461F,2012A&A...548A..56R}. {\rev The O\,I lines at 7773 and 8446\,\AA\ are also broad and could be related to accretion activity.} Two forbidden lines [O\,I]\,$\lambda$6300/6363 show very narrow line profiles with slightly blue-shifted peaks. They are proposed to be mainly due to prompt emission following UV photodissociation of OH molecules \citep{1998ApJ...502L..71S}. However, the atmosphere of Earth can also produce strong oxygen forbidden lines. Our observations may be contaminated by telluric emission. 

\onltab{4}{
\begin{table*}
\caption{EWs (\AA) of emission lines and Li absoprtion line of GW~Ori.\label{Tab:EW_line}}
\centering
\renewcommand{\tabcolsep}{0.03cm}
\begin{tabular}{cccccccccccccccc}
\hline\hline
ID  & H$\gamma$ &H$\beta$ & H$\alpha$ &He\,I & O\,I & O\,I & O\,I  &O\,I& Ca\,II &Ca\,II & Ca\,II & Ca\,II &Na\,I  &Li\,I\\
    & (4341\,\AA)     &(4861\,\AA) &(6563\,\AA) &(5876\,\AA) &(6300\,\AA) &(6363\,\AA) &(7773\,\AA) &(8446\,\AA) & (3933\,\AA) & (3968\,\AA)  &(8498\,\AA)&(8662\,\AA) &(5890/5896\,\AA) &(6708\,\AA)\\
\hline
1&-1.93&-4.63&  -30.29&-0.18&-0.28&-0.06&-0.29&-0.77&-10.21& -7.08&-6.09&-6.46&
-1.15& 0.20\\
2&-1.90&-4.33&  -28.93&-0.22&-0.29&-0.07&-0.37&-0.74& -8.86& -6.81&-5.29&-5.69&
-1.01& 0.20\\
3&-1.72&-3.73&  -24.21&-0.25&-0.29&-0.07&-0.27&-0.61& -8.45& -6.34&-5.00&-5.36&
-0.75& 0.21\\
4&-1.26&-2.96&  -20.27&-0.09&-0.26&-0.07&-0.01&-0.35& -6.55& -4.71&-3.12&-3.37&
-0.28& 0.23\\
5&-1.22&-3.00&  -20.50&-0.10&-0.27&-0.06&-0.11&-0.40& -6.57& -4.77&-3.40&-3.66&
-0.43& 0.23\\
6&-1.24&-3.01&  -20.28&-0.11&-0.26&-0.06&-0.10&-0.41& -6.55& -4.82&-3.31&-3.61&
-0.48& 0.22\\
7&-1.43&-3.45&  -22.99&-0.13&-0.27&-0.06&-0.14&-0.50& -7.46& -4.99&-3.87&-4.14&
-0.60& 0.23\\
8&-1.56&-3.66&  -23.55&-0.15&-0.27&-0.06&-0.13&-0.44& -7.06& -5.10&-3.60&-3.88&
-0.57& 0.22\\
9&-1.55&-3.54&  -23.34&-0.23&-0.26&-0.06&-0.13&-0.51& -7.84& -5.35&-4.00&-4.30&
-0.56& 0.22\\
10&-1.26&-3.12&  -22.21&-0.08&-0.26&-0.05&-0.06&-0.43& -6.48& -4.53&-3.21&-3.60&
-0.50& 0.23\\
11&-1.22&-2.92&  -20.61&-0.08&-0.25&-0.06&\nodata&\nodata& -7.77& -5.16&\nodata&
\nodata&-0.36& 0.20\\
12&-1.45&-3.36&  -22.28&-0.14&-0.25&-0.06&\nodata&\nodata& -8.06& -5.53&\nodata&
\nodata&-0.52& 0.22\\
13&-1.43&-3.38&  -22.41&-0.16&-0.25&-0.06&\nodata&\nodata& -8.19& -5.42&\nodata&
\nodata&-0.40& 0.21\\
14&-1.26&-2.93&  -19.07&-0.09&-0.27&-0.07&-0.00&-0.33& -6.54& -4.87&-3.26&-3.50&
-0.37& 0.22\\
15&-1.24&-2.85&  -18.89&-0.07&-0.27&-0.07&-0.03&-0.36& -6.87& -4.74&-3.27&-3.53&
-0.46& 0.22\\
16&-1.20&-2.73&  -18.01&-0.06&-0.26&-0.07&-0.04&-0.39& -6.15& -4.47&-2.97&-3.27&
-0.41& 0.22\\
17&-1.12&-2.64&  -18.00&-0.07&-0.27&-0.07&-0.03&-0.36& -6.03& -4.65&-2.93&-3.22&
-0.44& 0.22\\
18&-1.24&-2.81&  -19.35&-0.14&-0.27&-0.07&-0.10&-0.39& -6.27& -4.37&-3.08&-3.29&
-0.49& 0.22\\
19&-1.18&-2.85&  -19.04&-0.15&-0.27&-0.06&-0.09&-0.31& -6.28& -4.63&-3.09&-3.28&
-0.48& 0.22\\
20&-1.20&-2.84&  -18.69&-0.09&-0.28&-0.06&-0.04&-0.37& -6.92& -4.79&-3.30&-3.46&
-0.52& 0.22\\
21&-1.30&-3.01&  -19.35&-0.08&-0.27&-0.06&-0.04&-0.35& -7.05& -4.78&-3.26&-3.51&
-0.50& 0.23\\
22&-1.19&-2.86&  -18.65&-0.13&-0.26&-0.08&-0.03&-0.33& -6.48& -4.69&-2.97&-3.22&
-0.50& 0.22\\
23&-1.25&-2.88&  -18.84&-0.14&-0.27&-0.06&-0.05&-0.32& -6.56& -4.73&-3.03&-3.33&
-0.50& 0.22\\
24&-1.48&-3.17&  -20.43&-0.19&-0.27&-0.07&-0.12&-0.41& -7.00& -4.76&-3.31&-3.58&
-0.50& 0.23\\
25&-1.44&-3.20&  -20.57&-0.19&-0.27&-0.07&-0.16&-0.39& -7.04& -4.97&-3.36&-3.60&
-0.51& 0.22\\
26&-1.15&-2.68&  -19.15&-0.09&-0.27&-0.07&-0.01&-0.35& -6.21& -4.54&-2.98&-3.17&
-0.47& 0.23\\
27&-1.17&-2.73&  -19.06&-0.07&-0.26&-0.07&-0.01&-0.33& -6.16& -4.69&-3.01&-3.22&
-0.47& 0.24\\
28&-1.93&-4.21&  -24.20&-0.22&-0.26&-0.07&-0.21&-0.65& -8.44& -5.92&-4.39&-4.67&
-0.70& 0.22\\
29&-1.84&-4.04&  -23.82&-0.22&-0.27&-0.07&-0.19&-0.53& -7.93& -6.00&-4.32&-4.62&
-0.77& 0.22\\
30&-1.52&-3.44&  -21.68&-0.19&-0.26&-0.07&-0.10&-0.50& -7.37& -5.17&-3.59&-3.96&
-0.53& 0.22\\
31&-1.42&-3.30&  -21.36&-0.18&-0.27&-0.08&-0.10&-0.42& -7.08& -5.02&-3.45&-3.78&
-0.52& 0.23\\
32&-1.41&-3.23&  -21.43&-0.10&-0.28&-0.07&-0.08&-0.44& -6.66& -5.07&-3.40&-3.73&
-0.52& 0.23\\
33&-1.69&-3.53&  -22.86&-0.28&-0.30&-0.07&\nodata&\nodata& -8.32& -5.85&\nodata&
\nodata&-0.75& 0.24\\
34&-1.01&-2.25&  -19.46&-0.18&-0.32&-0.07&\nodata&\nodata& -8.08& -5.14&\nodata&
\nodata&-0.60& 0.21\\
35&-1.52&-3.41&  -21.98&-0.24&-0.30&-0.09&\nodata&\nodata& -8.70& -5.60&\nodata&
\nodata&-0.54& 0.23\\
36&-1.02&-2.31&  -19.10&-0.17&-0.34&-0.07&\nodata&\nodata& -8.02& -5.21&\nodata&
\nodata&-0.42& 0.21\\
37&-1.16&-3.21&  -19.67&-0.08&-0.31&-0.07&\nodata&\nodata& -6.53& -4.41&\nodata&
\nodata&-0.45& 0.23\\
38&-1.25&-3.24&  -21.94&-0.09&-0.27&-0.07&\nodata&\nodata& -7.54& -4.90&\nodata&
\nodata&-0.53& 0.21\\
39&-1.45&-3.47&  -23.33&-0.18&-0.30&-0.07&\nodata&\nodata& -7.47& -4.90&\nodata&
\nodata&-0.55& 0.23\\
40&-1.32&-3.35&  -22.39&-0.13&-0.30&-0.08&\nodata&\nodata& -7.74& -5.18&\nodata&
\nodata&-0.59& 0.22\\
41&-1.79&-4.28&  -28.70&-0.15&-0.34&-0.11&-0.27&-0.69& -6.86& -5.20&-4.76&-5.15&
-0.92& 0.22\\
42&-1.39&-4.41&  -31.58&-0.13&-0.37&-0.21&-0.02&-0.40&\nodata& -1.70&-4.00&-3.98
&-0.52& 0.24\\
43&-1.83&-4.32&  -31.13&-0.17&-0.34&-0.11&-0.16&-0.64& -7.80& -5.85&-4.34&-4.64&
-0.81& 0.22\\
44&-1.47&-3.46&  -27.38&-0.08&-0.35&-0.09& 0.03&-0.45& -6.22& -3.78&-3.07&-3.42&
-0.50& 0.21\\
45&-1.75&-4.47&  -30.98&-0.14&-0.35&-0.11&-0.13&-0.63& -8.07& -5.54&-4.85&-4.92&
-0.76& 0.22\\
46&-1.63&-4.02&  -27.78&-0.11&-0.36&-0.11& 0.03&-0.46& -7.32& -5.42&-3.69&-3.88&
-0.56& 0.23\\
47&-1.71&-4.19&  -27.03&-0.17&-0.35&-0.10&-0.07&-0.54& -7.88& -5.30&-4.15&-4.29&
-0.70& 0.23\\
48&-1.40&-3.65&  -26.44&-0.07&-0.35&-0.09&-0.07&-0.49& -6.95& -5.10&-3.55&-3.80&
-0.68& 0.22\\
49&-1.29&-3.07&  -23.84&-0.09&-0.34&-0.09&-0.11&-0.48& -6.47& -4.95&-3.59&-3.83&
-0.48& 0.23\\
50&-1.25&-2.99&  -22.36&-0.06&-0.33&-0.09&-0.06&-0.43& -5.89& -4.40&-3.28&-3.43&
-0.43& 0.23\\
51&-1.67&-3.88&  -25.78&-0.21&-0.33&-0.07&-0.16&-0.51& -7.25& -5.27&-3.44&-3.84&
-0.51& 0.21\\
52&-1.71&-3.90&  -25.54&-0.17&-0.33&-0.11&-0.13&-0.46& -7.00& -5.29&-3.47&-3.90&
-0.49& 0.23\\
53&-1.54&-3.55&  -25.24&-0.11&-0.32&-0.09&-0.05&-0.42& -6.42& -4.57&-3.19&-3.48&
-0.45& 0.24\\
54&-1.18&-2.80&  -22.46&-0.05&-0.32&-0.09&-0.03&-0.38& -5.74& -4.22&-2.85&-3.16&
-0.37& 0.23\\
55&-1.42&-3.55&  -27.59&-0.11&-0.32&-0.11& 0.12&-0.42& -6.15& -4.98&-3.73&-3.59&
-0.58& 0.23\\
56&-1.32&-3.37&  -26.22&-0.08&-0.32&-0.09&-0.03&-0.39& -6.98& -4.80&-3.35&-3.66&
-0.47& 0.24\\
57&-1.23&-3.02&  -23.09&-0.02&-0.29&-0.09&-0.04&-0.39& -5.31& -4.21&-3.15&-3.43&
-0.47& 0.24\\
58&-1.28&-3.08&  -23.15&-0.05&-0.32&-0.11&-0.01&-0.36& -5.83& -4.57&-3.37&-3.48&
-0.48& 0.23\\
\hline
\end{tabular}
\end{table*}
}

\subsection{Accretion}\label{Sec:accretion}
\subsubsection{Accretion-related emission lines}\label{Sec:accretion_line}

\begin{figure}
\centering
\includegraphics[width=1.\columnwidth]{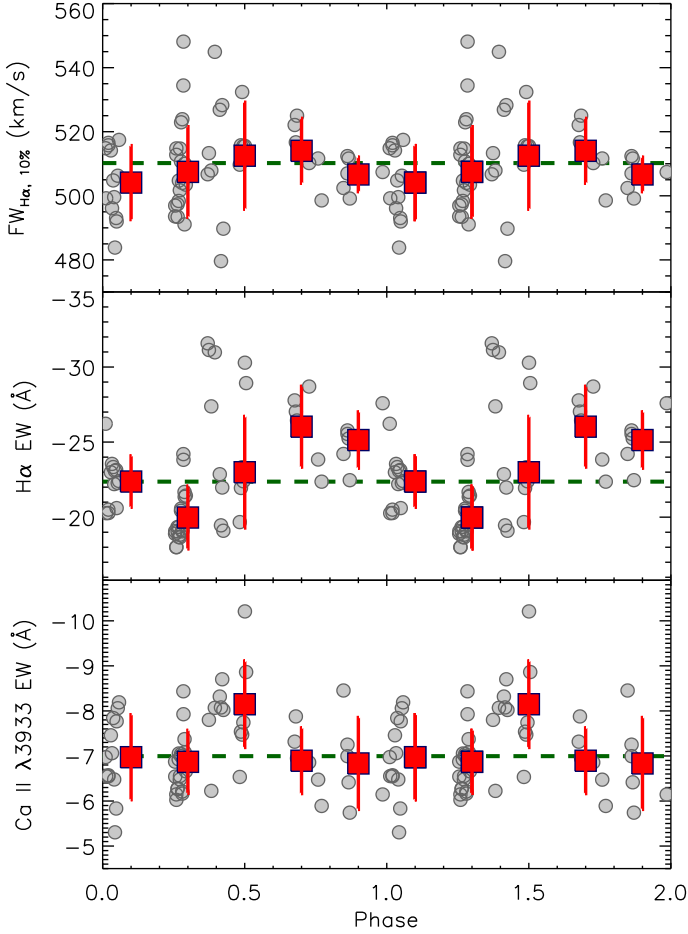}
\caption{$FW_{H\alpha, 10\%}$ values for the broad components from the H$\alpha$ line decomposition, and  EWs of H$\alpha$ and  Ca\,II\,$\lambda$3933 with respect to the orbital phases of GW~Ori~B. We divide the orbital phases into different bins and show the mean values in individual bins with filled squares, as well their standard deviations with error bars. The dashed line in each panel is the median value.}\label{Fig:Line_variation}
\end{figure}

The GW~Ori system has a low eccentricity ($e$=0.18) and, thus, is expected to show a smooth enhancement in the accretion rates with the orbital phases \citep{1996ApJ...467L..77A}. In Fig.~\ref{Fig:Line_variation}, we show the $FW_{H\alpha, 10\%}$ for the broad components (produced in accretion processes, see \sect~\ref{Sec:Halpha_Hbeta_lines}(d)) from the line decomposition, the EWs of the H$\alpha$ line, and the EWs of Ca\,II\,$\lambda$3933 line with respect to the orbital phases of GW~Ori~B. Both lines are good tracers of accretion. Note that our observations are not uniformly sampled over the orbital phases, which makes it hard to draw any definite conclusions. However, some hints can be seen in Fig.~\ref{Fig:Line_variation}. The $FW_{\rm H\alpha, 10\%}$ values are distributed around 510\,\kms\ over all the orbital phases. At some time we see that $FW_{\rm H\alpha, 10\%}$ reachs 550\,\kms at the  orbital phases around $\sim$0.4, corresponding to an increase of accretion rates by a factor of 2--3. The EWs of H$\alpha$ and Ca\,II\,$\lambda$3933 lines are scattered around $-$23 and $-$5\AA, respectively over all the orbital phases. {\rev We note sometimes the EWs of both lines increase by a factor of $\sim$1.5 at the orbital phases around $\sim$0.5, suggesting an increase of accretion rate by a factor of $\sim$2, according to the relations between the line luminosity of H$\alpha$ and  Ca\,II\,$\lambda$3933  to the accretion luminosity \citep{2009A&A...504..461F,2008ApJ...681..594H}.}

\begin{figure}
\centering
\includegraphics[width=1.\columnwidth]{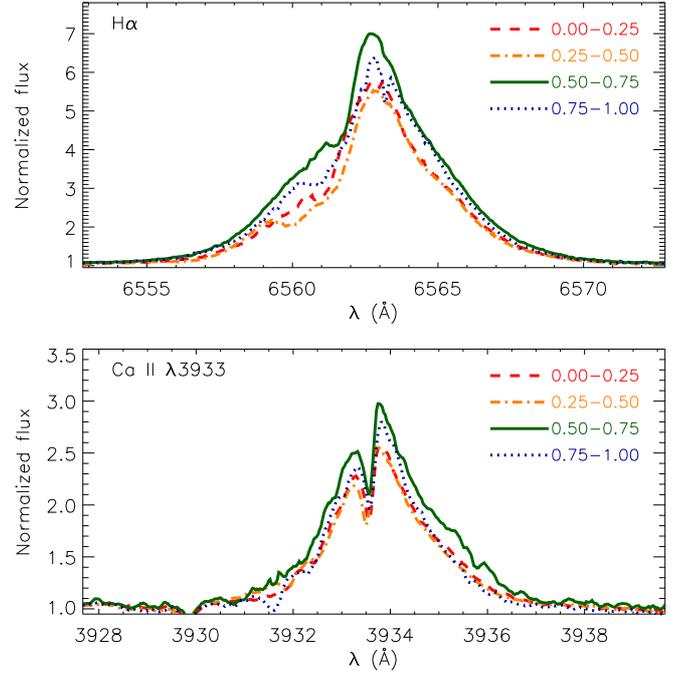}
\caption{The median line profiles of H$\alpha$ and Ca\,II\,$\lambda$3933. The line profiles are  within four phase bins, 0--0.25, 0.25--0.5, 0.5--0.75, and 0.75--1.0, and are shown with the dashed lines, dash-dotted lines, solid lines, and dotted lines, respectively.}\label{Fig:average_line}
\end{figure}

{\rev We divide the  $FW_{\rm H\alpha, 10\%}$ and EW values into different phase bins and show their mean values and the standard deviations in Fig.~\ref{Fig:Line_variation}. The mean $FW_{\rm H\alpha, 10\%}$  is quite constant over all the phase bins with a slightly larger standard deviation at phase $\sim$0.5 due to the occasional enhancement of accretion at this bin. The mean EWs of H$\alpha$ line seems to change from bin to bin, but the variations are small with the lowest mean EW of $\sim-$20\,\AA\ at the phase bin$\sim$0.3 and the highest mean EW of $\sim-$26\,\AA\ at the phase bin$\sim$0.7. Similar to the  $FW_{\rm H\alpha, 10\%}$ the standard deviation of H$\alpha$ EWs at the phase bin$\sim$0.5 is larger than those of others. The mean EWs of the  Ca\,II\,$\lambda$3933 line are very constant within different phase bins, although a slight increase of the mean EW is noted at the phase bin$\sim$0.5.} In Fig.~\ref{Fig:average_line}, we show the median line profiles of H$\alpha$ and  Ca\,II\,$\lambda$3968 within four bins of orbital phases, 0--0.25, 0.25--0.5, 0.5--0.75, and 0.75--1.0. For both H$\alpha$ and  Ca\,II\,$\lambda$3968,  we see the line profiles within the phase bins of 0--0.25, 0.25--0.5, and 0.75--1.0 are quite similar, and the line profiles within the phase bin of 0.5--0.75 are stronger than others.% which is attributed to the accidental increase of accretion rates within the bin. 

{\rev We have examined other accretion-related emission lines, such as H$\beta$, H$\gamma$, Na\,I\,D, and other Ca\,II lines. These lines exhibit similar behaviors to those of H$\alpha$ and Ca\,II\,$\lambda$3933, where the EWs of these lines are scattered around constant values over all the orbital phases of GW~Ori~B with a probable accidental increase of accretion at the phases around 0.5.  Since our observations do not cover the phases before and after the high accretion value at the same orbital period and, thus, we do not know the accretion rates before and after phase 0.5, it is not possible to establish whether the enhancement of accretion at the phases$\sim$0.5 is due to the orbital modulation or to an unrelated, more extended episode of increased accretion.} 

Here, we use the EWs of accretion-related emission lines as proxy of accretion of GW~Ori. However, the variations of the line EWs could be also attributed to variable brightness of a star. Long-term monitoring of GW~Ori with imaging has revealed that GW~Ori is a variable source \citep{2007A&A...461..183G}. Furthermore, the strength of some accretion-related emission lines, such as H$\alpha$, H$\beta$,  Ca\,II lines, and etc., may be contaminated by the stellar and disk winds, or be modulated by the stellar rotation (see Figs.~\ref{Fig:residual1} and~\ref{Fig:residual2}) if the accretion columns are not uniformly distributed on the stellar surface \citep{2013MNRAS.431.2673K}. These factors may in part produce the scatter seen in Fig.~\ref{Fig:Line_variation}.

\subsubsection{$U$-band excess}\label{Sec:uband-excess}

\begin{figure}
\centering
\includegraphics[width=1.\columnwidth]{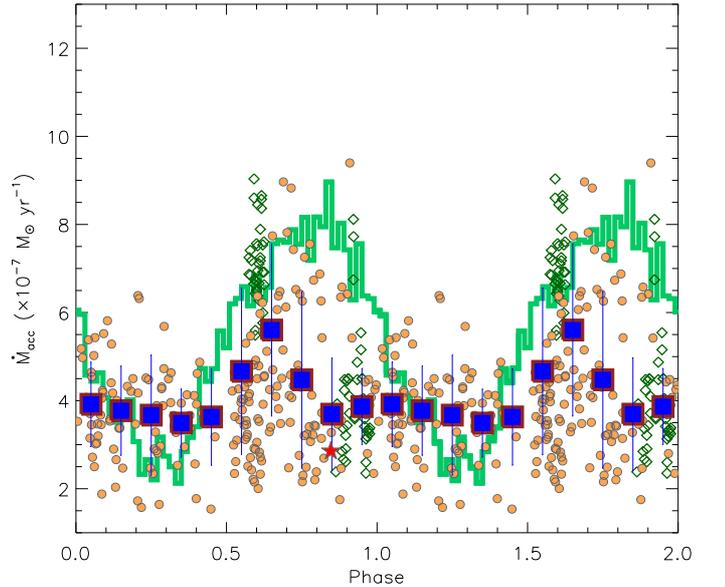}
\caption{The accretion rates derived from $U$-band excesses of GW~Ori, as plotted over the orbital phases of GW~Ori~B. The filled circles show the accretion rates estimated from the data in \citet{2007A&A...461..183G}; the open diamonds are for the accretion rates derived from the data in \citet{1988A&AS...75....1B}, and the star symbol mark the accretion rate from \citet{2004AJ....128.1294C}. We divide the orbital phases into different bins and show the mean value of accretion rates in each bin with a filled square, as well as its standard deviation with an error bar. The solid line show theoretical prediction for the total accretion rate of a close binary system in the low-eccentricity case from \citet{1996ApJ...467L..77A}. The theoretical result have been arbitrarily scaled to compare with the observations.}\label{Fig:Uexcess}
\end{figure}

The $U$-band excess is an excellent proxy for accretion rates and has been wide used in the literature \citep{1998ApJ...492..323G,1998ApJ...495..385H,2010ApJ...710..597S,2011A&A...535A..99M,2011A&A...525A..47R}. The object GW~Ori has been simultaneously observed for 20 years (1983--2003) in $U$ and other optical bands \citep{1988A&AS...75....1B,2007A&A...461..183G}. With these data, which have not been used to study accretion  properties of GW~Ori in the literature, we can properly calculate the accretion rates  from $U$-band excess and study variability of accretion at a timescale of several ten years. We use the observed $V$$-$$R$ color to estimate the visual extinction. For the photometric data in the Johnson system from \citet{2007A&A...461..183G}, we first convert them to Cousins system by using the relation from \citet{1983AJ.....88..439L}: ($V$$-$$R$)$_{C}$=$-$0.0320+0.71652$\times$($V$$-$$R$)$_{J}$. The visual extinction is then estimated by comparing the observed ($V$$-$$R$)$_{C}$ color with the intrinsic ($V$$-$$R$)$_{C}$ color for synthetical spectra with $T_{\rm eff}$=5500\,K and log~$g$=3.5 in \citet{2013ApJS..208....9P} and by using an extinction law in Landolt $UVR$ bands from \citet{2011ApJ...737..103S} in the case of $R_{\rm V}$=3.1. We use the visual extinction to deredden the photometry. The $U$-band magnitude of the stellar photosphere is obtained from the dereddened V-band photometry and an intrinsic $U$$-$$V$ color for synthetical spectra with $T_{\rm eff}$=5500\,K and log~$g$=3.5 in \citet{2013ApJS..208....9P}. We convert the $U$-band photometry to $U$-band luminosity using the zero-point flux and bandwidth for $U$ band \citep[4.19$\times$10$^{-9}$\,erg\,s$^{-1}$\,cm$^{-2}$\AA$^{-1}$ and 680\,\AA, respectively;][]{2010ApJ...710..597S}. We derive the $U$-band excess emission by subtracting the photospheric emission from the dereddened emission of GW~Ori in $U$ band and then convert it to accretion luminosity according to the empirical relation in \citet{1998ApJ...492..323G}. The inferred accretion luminosities are then converted into mass accretion rates using the following relation:

\begin{equation}
\dot{M}_{\rm acc}=\frac{L_{{\rm acc}}R_{\star}}{{\rm G}M_{\star} (1-\frac{R_{\star}}{R_{\rm in}})},
\end{equation}

%treated, RvB
\noindent where $R_{\rm in}$ denotes the truncation radius of the disk, which is taken to be 5\,$R_{\star}$ \citep{1998ApJ...492..323G}. The parameter G is the gravitational constant, $M_{\star}$ and $R_{\star}$ is the stellar mass and radius of GW~Ori~A, respectively.

We find that GW~Ori shows excess emission in all collected $U$-band photometry, suggesting that GW~Ori has been accreting for 20 years. In total, we have 307 estimates of the accretion rate for GW~Ori during the period of 1983--2003. The typical uncertainty of measurement is $\sim$2$\times$10$^{-7}$\,\accunit. The mean accretion rate of GW~Ori is $\sim$4$\times$10$^{-7}$\,\accunit, which is consistent with the result (2.56--3.15$\pm$1.99$\times$10$^{-7}$\,\accunit) in \citet{2004AJ....128.1294C}. In Fig.~\ref{Fig:Uexcess}, we show the accretion rates of GW~Ori with respect to the orbital phases of GW~Ori~B. Similar to Fig.~\ref{Fig:Line_variation}, our data are not uniformly sampled over the orbital phases. Therefore, we would not make any conclusion from it. Instead, we only describe the hints from  Fig.~\ref{Fig:Line_variation}. We note that the accretion rates of GW~Ori are scattered around $\sim$4$\times$10$^{-7}$\,\accunit\ over all orbital phases, indicating that the accretion rate of GW~Ori is mostly constant. We also see that the accretion of GW~Ori are occasionally enhanced by a factor of $\sim$2--3 within the orbital phases of $\sim$0.6--0.8. We divide the data into different phase bins and calculate the mean values and the standard deviation of the measurements. The results are also shown in Fig.~\ref{Fig:Uexcess}. The mean values are generally constant within different phase bins. In addition, besides the phase bins around 0.6--0.8, the standard deviations of accretion rates are $\sim$1$\times$10$^{-7}$\,\accunit. Within the orbital phases of 0.6--0.8, the standard deviations are about twice the typical value within other phase bins, which is attributed to the enhancement of accretion at these phases. Similar to the accretion-related emission lines in \sect~\ref{Sec:accretion_line}, the $U$-band data also do not cover the phases before and after the enhancement of accretion at the same orbital period. {\subrev Therefore, there is no evidence that the increase of the accretion rates at phases 0.6--0.8 could be due to orbital modulation.}

\subsection{Spectral energy distribution}\label{Sec:SED}

\begin{figure}
\centering
\includegraphics[width=1.\columnwidth]{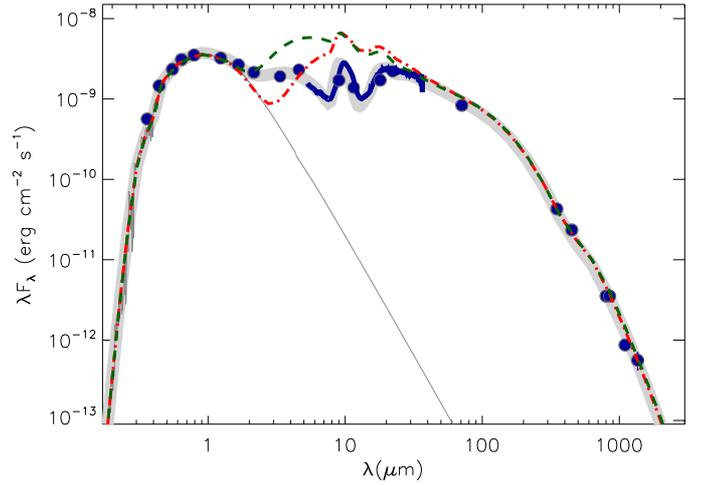}
\caption{The observed SED of GW~Ori. The broad band photometry is shown with the filled circles, and the IRS spectrum of this source is displayed in solid line. The thick gray line shows a best-fit model. The dashed line shows the model SED for a disk with  $M_{\rm disk}$=0.14\,\Msun, $H_{\rm out}$/$R_{\rm out}$=0.235, and $R_{\rm in}$=2.3\,AU, and the dot-dashed line display another model SED for a disk with  $M_{\rm disk}$=0.14\,\Msun, $H_{\rm out}$/$R_{\rm out}$=0.2, and $R_{\rm in}$=8\,AU. The photospheric emission level is indicated with a thin gray curve.\label{Fig:SED}}
\end{figure}

We construct the SED of GW~Ori by using the $UBVR_{\rm C}I_{\rm C}$ photometry from \citet{2004AJ....128.1294C}, the  $JHK_{\rm s}$ photometry from the 2MASS survey \citep{2006AJ....131.1163S}, the photometry at 3.4, 4.6, 12, and 22\,\mum\ from the WISE survey \citep{2010AJ....140.1868W}, the photometry at 9 and 18\,\mum\ from the AKARI survey \citep{2010A&A...514A...1I}, and the fluxes at 350, 450, 800, 850, 1100, 1360\,\mum\ from \citet{1995AJ....109.2655M}. We obtain the 5$-$37\,\mum\  low-resolution IRS spectrum\footnote{The extracted spectra are based on the droopres products processed through the S18.7.0 version of the Spitzer data pipeline. See the detail spectral extraction procedure is described in \citet{2008ApJ...683..479B}.}   and the MIPS 70\,\mum\ photometry\footnote{The aperture photometry of GW~Ori was performed with a 60$''$ aperture and a sky annuli of 71.5$''$ and 119.2$''$, which give a flux of 19.7$\pm$0.1\,Jy.} of GW~Ori from the Spitzer data archive (Program ID 40145 and 20339, respectively). We do not use the fluxes from the IRAS survey. A comparison of the fluxes at IRAS 60\,\mum\ and MIPS 70\,\mum\ show the IRAS 60\,\mum\ flux is $\sim$1.7 times of the flux at the MIPS 70\,\mum, suggesting that the IRAS observations with poor spatial resolutions are contaminated by the dust emission in the field near GW~Ori. Figure~\ref{Fig:SED} shows the SED of GW~Ori. The broad dip around the silicate 10\,\mum, as noted by \citet{1991AJ....101.2184M}, is clearly seen in the IRS spectrum. 

\begin{figure}
\centering
\includegraphics[width=1.\columnwidth]{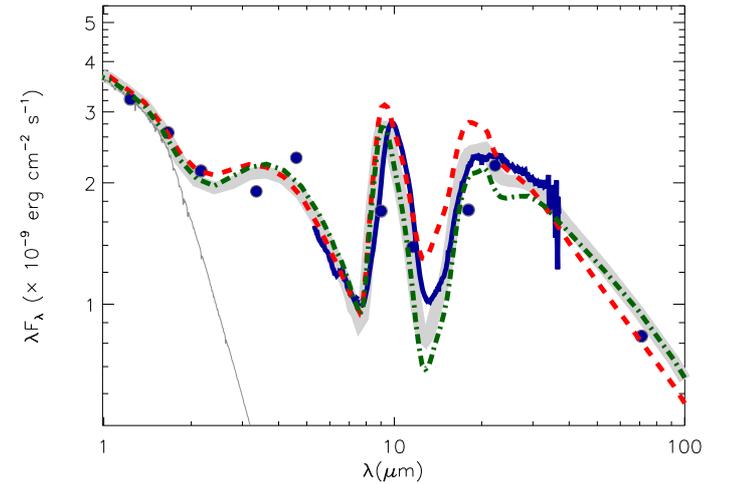}
\caption{The observed SED of GW~Ori within a wavelength range of 1--100\,\mum. The dashed line shows the model SED for a disk with $R_{\rm gap}$=25\,AU  and a dash-dotted line display another model SED for a disk with  $R_{\rm gap}$=55\,AU. Others in the figures are same as in Fig.~\ref{Fig:SED}. \label{Fig:in_SED}}
\end{figure}

\subsubsection{Modeling the SED of GW~Ori}\label{Sec:Model_SED1}

We use the 2D RADMC code from \citet{2004A&A...417..159D} to model the SED of GW~Ori.  The stellar parameters adopted in the models are $T_{\rm eff}$=5500\,K, $R_{\star}$=7.6\,$R_{\odot}$, and $M_{\star}$= 3.9\,\Msun. In the calculations, the inner disk radius ($R_{\rm in}$) is left as a free parameter with the disk mass ($M_{\rm disk}$). The outer disk radius ($R_{\rm out}$) is fixed to be $\sim$500\,AU \citep{1995AJ....109.2655M}. We assume a pressure scale height ($H_{\rm P}$) that varies as a power law with the disk radius (R), $H_{\rm P}/R=R^{1/7}$. The scale height at the outer disk radius is parameterized as $H_{\rm out}$/$R_{\rm out}$ and set to be a free parameter. The disk surface density ($\Sigma$) is estimated from $M_{\rm disk}$, assuming a distribution $\Sigma\propto R^{\alpha}$ and setting $\alpha$=$-$1. We take a gas-to-dust ratio to be 100 and use a power-law size distribution with an exponent of $-3.5$ for the sizes of dust with a minimum size of 0.1\,\mum\ and a maximum grain size of 1000\,\mum. Two populations of amorphous dust grains (25\% carbon and 75\% silicate) are included in the calculations. We assume that the disk axis is aligned with the stellar rotation axis of GW~Ori (see \sect~\ref{sect:rv}) and use a moderate disk inclination of $\sim$40$^\circ$ to calculate the SEDs from disk models. We vary the free parameters in disk models to calculate the SEDs. In Fig.~\ref{Fig:SED}, we show one example model SED calculated with $M_{\rm disk}$=0.14\,\Msun, $H_{\rm out}$/$R_{\rm out}$=0.235, and $R_{\rm in}$=2.3\,AU. The model SED has been reddened with $A_{\rm V}$$\sim$1.4 to fit the photometry of GW~Ori in optical bands\footnote{Hereafter, all the calculated SEDs are reddened to fit the SEDs of GW~Ori in optical bands before we compare them with the observations.}. We note that the model SED can fit the observed SED at wavelengths $\gtrsim$30\,\mum\ very well but exhibits much stronger excess emission at infrared wavelengths of 2--30\,\mum\ than the observations, indicating that the $R_{\rm in}$ of the disk model is too small. We increase $R_{\rm in}$ to 8\,AU and set $H_{\rm out}$/$R_{\rm out}$=0.2 and $M_{\rm disk}$=0.14\,\Msun. The calculated SED is shown in Fig.~\ref{Fig:SED}. Similar to the previous model, the new SED can fit the observed SED at wavelengths $\gtrsim$30\,\mum\ but shows much stronger excess emissions at a wavelength range of 6--30\,\mum\ compared with the observations. We also note that the new disk model exhibits no excess emission at wavelengths $\lesssim$3\,\mum, which is inconsistent with the observations. The above two experiments give a hint that it is hard to reproduce the SED of GW~Ori using typical disk models.

Since GW~Ori is a triple system with a second companion at a projected distance $\sim$8\,AU, we would expect a gap in the disk  created by the companions, as shown in the simulations \citep{1994ApJ...421..651A,1996ApJ...467L..77A,2008MNRAS.391..815P}. Such disk models with gaps can have moderate excess emission at near- and mid-infrared wavelengths by heating the small amount of dust particles in the gap. Furthermore, the strong and narrow silicate emission feature at 10\,\mum\ shown on the IRS spectrum of GW~Ori indicates that the dust in the gap is dominated by a small amount of tiny dust particles. To include a gap in the disk models, we add a new parameter ($R_{\rm gap}$) as the gap size. In the gap, the dust also consists of the two populations of amorphous dust grains with the same power-law distribution of sizes as in the outer disk but with a minimum size of 0.005\,\mum\ and a maximum size of 1\,\mum. In the calculations, the model grid is refined at the edge of the gap to {\subrev improve resolution in a place where opacity changes.}

The distribution of dust in the gap is parameterized like a small independent disk. The outer radius ($R_{\rm gap}$), the inner disk radius ($R_{\rm in, gap}$), the total mass ($M_{\rm in, gap}$), and the power exponent ($\alpha_{\rm in}$), which assumes a distribution $\Sigma_{\rm in, dust}\propto R^{\alpha_{\rm in}}$ for the dust surface density ($\Sigma_{\rm in, dust}$)  in the gap, are all left as free parameters. Hereafter, we refer to the {\it inner disk} and {\it outer disk} as the one in the gap and the one with radius$\geq$$R_{\rm gap}$, respectively.  We vary the free parameters in the  models to search for the best-fit models. We find that the SED of GW~Ori can be fitted very well with a model with $R_{\rm gap}$=45\,AU, $R_{\rm in, gap}$=1.8\,AU, and a total disk mass $\sim$0.21\,\Msun. The best-fit SED is shown in Fig.~\ref{Fig:SED}. The parameters for the best-fit model are shown in Table~\ref{Tab:disk_par} (Type~1).

We explore the parameter space of $R_{\rm gap}$ by varying $R_{\rm gap}$ and the other free parameters to fit the SED of GW~Ori. In Fig.~\ref{Fig:in_SED}, we show the SEDs that are calculated from disk models with  $R_{\rm gap}$=25 and 55\,AU, respectively. For clarity, we only show the SEDs at the wavelength range where different models are distinguishable from each other. A disk model with  $R_{\rm gap}$$\lesssim$25\,AU would have a stronger excess at 13-20\,\mum\ than the observations, while one with $R_{\rm gap}$$\gtrsim$55\,AU would produce weaker excess emission. Thus, we set $R_{\rm gap}$=25--55\,AU and take 45\,AU as the best value of $R_{\rm gap}$. However, we must stress that our SED fits cannot provide strong constraints on $R_{\rm gap}$, as well as other parameters, since the disk models have too many free parameters. {\rev The gap size would be affected by assuming different dust compositions and inner wall sizes. It is also affected if it is not pure gap but filled with non-axisymmetric structures caused by the companions.} However, our attempt to fit the SED of GW~Ori provides an important hint that disk models with gaps can reproduce the SED of GW~Ori very well.

{\subrev In our SED modeling, we only include two ingredients, the primary GW~Ori~A and a disk surrounding GW~Ori. Recently, \citet{2011A&A...529L...1B} find that the brightness of GW~Ori~B can be $\sim$60\% of the one of GW~Ori~A. Since GW~Ori~A and B are very close compared to the gap and the luminosity of GW~Ori~B has been added to that of GW~Ori~A in the modeling, it would not matter if we include one or two illuminating sources at the disk center. However, due to the interaction between the disk and the stellar system, the disk is most likely non-axisymmetric. The GW~Ori~C is much less luminous than GW~Ori~A and B \citep{2011A&A...529L...1B}, but it could also locally heat the disk and break the disk symmetry.}

\begin{table}
\caption{Disk model parameters for GW~Ori.\label{Tab:disk_par}}
\centering
\begin{tabular}{ccccc}
\hline\hline
Parameters                   &Type\,1       &Type\,2       &Type\,3\\
\hline
\multicolumn{4}{c}{{\bf Central star}}\\
Effective temperature         &\multicolumn{3}{c}{5500\,K}\\
Radius                        &\multicolumn{3}{c}{7.5\,$R_{\odot}$}\\
Mass                          &\multicolumn{3}{c}{3.9\,$M_{\odot}$}\\
\hline
\multicolumn{4}{c}{{\bf Optically Thin Inner Disk}}\\
Inner radius (AU)                 &1.8        &0.80      &0.52   \\
%Surface density distribution\\
$\Sigma\propto R^{\alpha}$    &$-$2.4         &$-$0.2         &$-$9      \\
Total dust Mass (\Mearth)                   &6.7$\times$10$^{-4}$ &4.7$\times$10$^{-2}$    &6.0$\times$10$^{-5}$           \\
The scale height             &0.3           &0.24           &0.39     \\
\hline
\multicolumn{4}{c}{{\bf Outer disk}}\\
$R_{\rm gap}$                &\multicolumn{3}{c}{45\,AU}\\
Outer radius                &\multicolumn{3}{c}{500\,AU}\\ 
Total Mass (gas+dust)                 &\multicolumn{3}{c}{0.21\,\Msun}\\
The scale height            &\multicolumn{3}{c}{0.15}        \\
%Surface density distribution\\
$\Sigma\propto R^{\alpha}$   &\multicolumn{3}{c}{$-$1}      \\
\hline
\end{tabular}
\end{table}

\subsubsection{Variations in the SED of GW~Ori}\label{Sec:var_SED}
 
\begin{figure}
\centering
\includegraphics[width=1.\columnwidth]{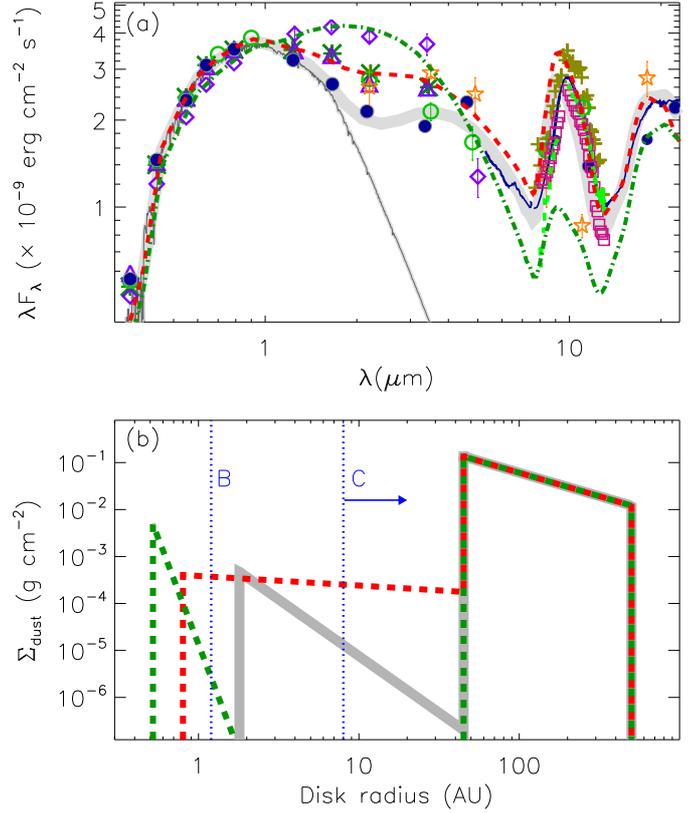}
\caption{(a) SEDs of GW~Ori constructed with the data observed at different epochs. Significant variations can be seen in the SED of GW~Ori at near-infrared bands. Open stars mark the observations obtained at November 1971 \citep{1973MNRAS.161...97C}. Open circles mark the data from November 1974 \citep{1976MNRAS.174..137C}. Asterisks display the data from December 1981 \citep{1983AJ.....88.1017R}. Open diamonds and triangles show the average photometry from December 1984 and December 1986, respectively \citep{1988A&AS...75....1B}.  Plus symbols around 10\,\mum\, show the data obtained on December 1978 \citep{1980MNRAS.191..499C}, and the dashed lines  around 10\,\mum\, display the data obtained on December 2002 \citep{2003A&A...412L..43P}. The open boxes around 10\,\mum\, show the data observed on March 2005 \citep{2009A&A...502..367S}. Other symbols are the same as shown in \fig~\ref{Fig:SED}. The thick gray curve shows calculated SED from Type\,1; the dashed line for the SED from Type\,2, and the dash-dotted line for the SED from Type\,3. (b) Surface density profiles for the dust in three model disks shown in panel~(a). The thick gray lines are for Type\,1; the dashed lines are for Type\,2, and the dash-dotted lines are for Type\,3. The locations of two companions in GW~Ori are also marked. The right-pointing arrow means that  the separation for  from GW~Ori~C to the primary is a lower limit. \label{Fig:var_SED}}
\end{figure}

\citet{1995AJ....109.2655M} constructed the SED of GW~Ori using the data from the literature \citep{1983AJ.....88.1017R,1976MNRAS.174..137C,1980MNRAS.191..499C,1973MNRAS.161...97C}, the IRAS survey, and their own observations at submillimeter and millimeter. The main difference between our SED and theirs is in the infrared bands. We use the infrared data from the 2MASS survey, WISE survey, and the Spitzer observations. In Fig.~\ref{Fig:SED}, we compare two SEDs at wavelengths $\lesssim$20\,\mum. We note that both SEDs are generally consistent with each other except in the wavelength range between the wavelengths of 1.6--3.5\,\mum. At 1.6--3.5\,\mum, the Mathieu's SED  shows  stronger excess emission than ours. In Fig.~\ref{Fig:var_SED}, we also show the observations of GW~Ori collected from other works \citep{1988A&AS...75....1B,2003A&A...412L..43P,2009A&A...502..367S}. In \citet{1988A&AS...75....1B}, they monitored GW~Ori in multiple bands during December 1984 and and December 1986. In Fig.~\ref{Fig:var_SED}, we show the average photometry and the standard deviation in each band during each period. We note the data obtained in December 1986 show similar excess emissions at 1.6--3.5\,\mum\ to those of \citet{1995AJ....109.2655M}. Furthermore, the data obtained in December 1984 show strongest excess emissions at 1.6--3.5\,\mum\ among all the observations. During the same period, the observation around 5\,\mum\ show a dramatically decrease in flux and exhibits lowest excess among all the data. These temporal variations in near-infrared bands  cannot be simply explained as variable extinctions, since their simultaneous data in optical bands do not show a significant difference. One promising explanation could be the change of inner-disk structure, since the fluxes of GW~Ori at 1.6--3.5\,\mum\ are dominated by the emission from the inner-disk region. 

In Fig.~\ref{Fig:var_SED}, the SEDs of GW~Ori in near- and mid-infrared bands can be classified into three types.  The type~1 SED is referred to our constructed SED, the type~2 is for the Mathieu's, and the type~3 is the one observed in December 1984 from \citet{1988A&AS...75....1B}. The detail modeling of the type~1 SED is described in \sect~\ref{Sec:Model_SED1}, and their best-fit parameters are listed in Table~\ref{Tab:disk_par}. Here, we model the other two types. Given that we only have few multi-epoch observations at longer wavelengths ($\gtrsim$10\mum) and these data  show no obvious variations, we fix the parameters for the outer disk in the modeling and use the best-fit parameters for the outer disk from fitting type~1 SED. We vary the parameters for the inner disk (see \sect~\ref{Sec:Model_SED1} ) to reproduce the other two types of SEDs.

Using the above simple models, we find that all the observed SEDs of GW~Ori can be naturally explained by the change of the structure of the inner disk. In Fig.~\ref{Fig:var_SED}, we show the best-fit model SEDs for the three types of SEDs and the distribution of the surface densities for the disk models. The best-fit parameters  are listed in Table~\ref{Tab:disk_par}. We note that a small amount of dust ($\sim$6$\times 10^{-5}$--5$\times 10^{-2}$\,\Mearth) in the gap is enough to produce the observed excess emission at near- and mid-infrared bands. In each model the dust surface densities in the gap are reduced by several orders of magnitude, as compared with those at $R_{\rm gap}$. However it is unknown whether the gas surface densities in the gap decrease by a similar factor to those of dust, since our SED modeling cannot provide any constraints on the gas material in the disk. From the type~1 to type~3 SED, the increase of excess emission in near-infrared bands is interpreted as the decrease of $R_{\rm in}$. The type~1 and type~2 models have similar excess emissions at wavelengths$\gtrsim$5\,\mum, while the type~3 model produces an abrupt drop of fluxes at wavelengths$\sim$5\,\mum\ due to the dust material accumulating at radii $\sim$0.5\,AU. Furthermore, the type~3 model produces a weaker silicate emission feature at 10\,\mum\ compared with the other two models. Unfortunately, there are no simultaneous data around 10\,\mum\ available, which makes this hypothetical disk structure hard to be tested.

Here, we simply assume that the dust in the gap is distributed in a disk structure. In reality, the distribution of dust material in the gap could be very complex due to the existence of two companions and the dust filtering in GW~Ori \citep{1996ApJ...467L..77A,2002A&A...387..550G,2006MNRAS.373.1619R,2012ApJ...755....6Z}. In a binary system, simulations show that the material in the gap is mainly distributed in a set of two gas streams, which penetrate the gap and transfer the gas material from circumbinary disk to the central binary \citep{1996ApJ...467L..77A,2002A&A...387..550G}. In the simulations, the gas streams only account for $\sim$10\% of the gap area and can carry at most 10$^{-3}$ of the mass that would exist inside the gap for an unperturbed disk \citep{1996ApJ...467L..77A}. The dynamic evolution of the gas streams can explain the variation in the SED. For GW~Ori, a second companion does complicate this two-stream scenario. {\rev When including the effect of  dust filtering, which can prevent large dust particles from crossing the gap in the disk, the scenario could be more complex.} New simulations specified for GW~Ori are extremely useful for understanding this system.

\begin{figure}
\centering
\includegraphics[width=1.\columnwidth]{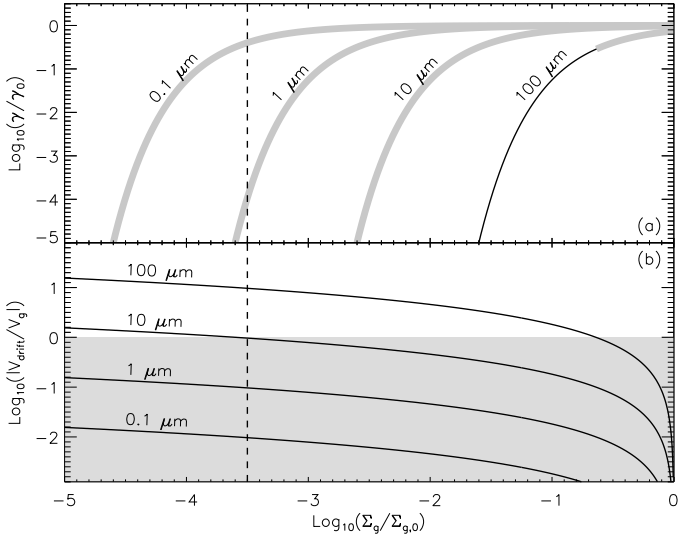}
\caption{(a) Dust depletion factor due to the dust filtration by gap edges as a function of the gap depth for 4 grain sizes of dust, which include 0.1, 1, 10, and 100\,\mum. The dust depletion factor is calculated assuming that the dust diffusion balances the dust drift. For each grain size, the thick gray line shows the regime where the gas velocity amplification at the gap edge dominates (see panel~b). In this case, the dust depletion factor is only the lower limit at the gap depth. The dashed line marks the gap depth at which the dust particle with size $\geq$10\,\mum\ can be trapped because the dust drift velocity is larger than the gas velocity (see panel~b). (b) The ratio of the dust drift velocity to the gas velocity with respect to the gap depth for the dust grains with different sizes. In the calculation, we consider the amplification of gas velocity in the gap and neglect the effect of dust diffusion. The gray-filled region marks, where the ratio is smaller than 1 and dust particles can pass through the gap. The dust grains will be prevented from crossing the gap when the ratio is larger than 1. The dashed line is same as in panel~(a). The figure is made using the equations in \citet{2012ApJ...755....6Z}. \label{Fig:Dust}}
\end{figure}

\section{Discussion: Disk evolution in binary/multiple systems}\label{Sec:discussion}

\subsection{Accretion}

In December 1984 and December 1986, \citet{1988A&AS...75....1B} observed GW~Ori in multiple bands. {\subrev These data shows an obvious change} in the SED of GW~Ori in near-infrared bands (see \sect~\ref{Sec:var_SED} and Fig.~\ref{Fig:in_SED}). We explain the SED variations as the change of the dust distribution in the gap and find that the dust mass in the gap in December 1984 could be reduced by three orders of magnitude compared with the one in December 1986.{\newrev From the $U$-band photometric data from \citet{1988A&AS...75....1B}, the accretion rates of GW~Ori at the two epochs are similar ($\sim$4$\times$10$^{-7}$\,\accunit), However, the mean accretion rate of GW~Ori in February 1984 is $\sim$7$\times$10$^{-7}$\,\accunit, which is about  two times larger than the typical value ($\sim$4$\times$10$^{-7}$\,\accunit, see \sect~\ref{Sec:uband-excess}). The latest observations of GW~Ori before February 1984 were acquired on December 3, 1981 by \citet{1983AJ.....88.1017R}, in which the $U$-band photometric data suggests an accretion rate $\sim$4$\times$10$^{-7}$\,\accunit. The enhancements of accretion for GW~Ori lasted for less than 3~years and could accrete the material of less than 0.7\,\Mearth. With an assumption of gas-to-dust ratio $\sim$100--1000, the total accreted mass of dust is $\sim7\times10^{-3}-7\times10^{-4}$\,\Mearth, which is comparable to the dust masses in the gap, as suggested by our SED modeling (see Table~\ref{Tab:disk_par}). Therefore, it is possible that enhanced accretion in GW~Ori can drain the material in the gap on a timescale of years, besides the gas-to-dust ratio in the gap is much larger than 1000.}

The simulations of \citet{1996ApJ...467L..77A} show that a binary system can have pulsed accretion modulated by the orbital motion. For the low-eccentricity case like GW~Ori, the theoretical works predict a smoothly enhanced accretion activity. Our observations show that the accretion rate of GW~Ori is mostly constant and presents an accidental increase at orbital phases of 0.5--0.8 by a factor of 2--3. In Fig.~\ref{Fig:Uexcess}, we compare the observations with the theoretical prediction. The simulation data are arbitrarily scaled to compare with the observations. We note the the theory can explain the observational data  in the orbital phases at which the enhanced accretion happens, as well as the level of the enhancement of accretion. However, as discussed in \sect~\ref{Sec:accretion}, we find the accretion rate of GW~Ori is mostly constant {\rev over all the orbital phases of GW~Ori~B}, which is inconsistent with the simulations from \citet{1996ApJ...467L..77A}. The simulations of binary systems also predict that low-mass companions accrete more masses than the primaries \citet{1996ApJ...467L..77A}. However, as shown in Fig.~\ref{Fig:residual1} and \ref{Fig:residual2}, the EWs of accretion-related emission lines are modulated by the rotation of GW~Ori~A. This finding suggests that the accreting masses are mostly funneled onto GW~Ori~A, which also contradicts with the results in \citet{1996ApJ...467L..77A}. {\subrev The spatial resolution of the simulations from \citet{1996ApJ...467L..77A} may be too low to resolve the  bridge-like stationary shock  between the accretion disks of the two stellar components in binary systems, which has been found in the new hydrodynamical simulations \citep{2011Ap&SS.335..125F,2012ARep...56..686B}. Due to this collision, most of the material in the circumsecondary disk can be accreted onto the primary, which leads to higher rate of accretion onto the primary than onto the secondary \citep{2011Ap&SS.335..125F}. Our result is consistent with the new simulations.}

\subsection{The disks in GW~Ori}

\subsubsection{Dusty circumstellar disks in GW~Ori}

 \citet{1991AJ....101.2184M,1995AJ....109.2655M} successfully reproduced the SED of GW~Ori using two-disk models with one circumstellar disk around GW~Ori~A and one circumbinary disk surrounding the GW~Ori system. In their best-fit model, the circumstellar disk around GW~Ori~A has an inner radius of 0.025\,AU and an outer radius of 0.17\,AU. The circumbinary disk starts from 3.3\,AU and extends to 110\,AU. Inside the gap between the circumstellar disk and the circumbinary disk, they add a small amounts of dust to reproduce the strong silicate emission feature at 10\,\mum. Our disk models are slightly different from the Mathieu's since we do not include a dusty circumstellar disk around GW~Ori~A. Our arguments are follows.

The circumstellar disk around each component in a binary can be externally truncated by the tidal interaction with the companion, and the size of circumstellar disk is related to the eccentricity, the semi-major axis, and the mass ratio of the binary system \citep{2005MNRAS.359..521P}. Assuming the orbital axis is aligned with the rotation axis of GW~Ori~A , the mass of GW~Ori~B is $\sim$0.3--0.7\,\Msun\ (see \sect~\ref{sect:rv}). According to the formula in \citet{2005MNRAS.359..521P}, the sizes of circumstellar disks around  GW~Ori~A and B would be $\sim$0.3, and 0.1\,AU, respectively. Furthermore, the inner radius of a dusty disk can be determined by the evaporation of dust \citep{2001ApJ...560..957D}. For GW~Ori~A with $T_{\rm eff}$=5500\,K and $L_{\star}$=48\,\Lsun, the inner radius of its dusty disk is estimated to be  $\sim$0.5\,AU, assuming a dust evaporation temperature ($T_{\rm evap}$) $\sim$1500\,K,  according to the formula in \citet{2001ApJ...560..957D}.  Therefore, it is very likely that there is no dusty disk around GW~Ori~A. The stellar parameters of GW~Ori~B is very uncertain. For a  0.3--0.7\,\Msun\ PMS star  at an age of $\sim$1\,Myr, its $T_{\rm eff}$ and  $L_{\star}$ is expected to be $\sim$3400--4000\,K and $\sim$0.7--1.7\,\Lsun, respectively \citep{2000A&A...358..593S}. With these stellar parameters and $T_{\rm evap}$=1500\,K, the inner radius is $\sim$0.07\,AU. If the orbital inclination of GW~Ori~B is much smaller ($\sim$10$^{\circ}$) as suggested in \citet{2011A&A...529L...1B}, the mass of GW~Ori~B could be 1.4\,\Msun\ with $T_{\rm eff}$$\sim$4500\,K and  $L_{\star}$$\sim$3.7\,\Lsun\ \citep{2000A&A...358..593S}, which gives an inner radius$\sim$0.13\,AU. Thus, we may not expect there is a dusty disk around GW~Ori~B, too. For GW~Ori~C, it is very hard to characterize its disk properties, given that its stellar parameters are unknown. Thus, we neglect dusty circumstellar disks  in our models. However, the gaseous circumstellar disk(s) must be harbored  by at least one stellar member in GW~Ori since GW~Ori is still accreting (see \sect~\ref{Sec:accretion}).

\subsubsection{Dust filtration}

To reproduce the strong and sharp silicate feature at 10\,\mum, a small population of tiny dust particles needs to be included in the gap, which would be consistent with a dust filtration scenario in GW~Ori. This effect has been proposed to explain the SEDs of transitional disk objects (TOs) and is supposed to occur in a disk with a gap where the  gaseous pressure gradient can only let the dust grains be smaller than some critical size through the gap and trap the large dust particles, because of the dust radial drift \citep{2006MNRAS.373.1619R}. Recently, a direct evidence of dust trapping has been found in Oph~IRS~48 with ALMA \citep{2013Sci...340.1199V}.

In \citet{2012ApJ...755....6Z}, they discuss two effects to resist the dust radial drift: dust diffusion and amplified gas radial velocity at the gap edge. When the dust diffusion is against the dust drift without regard for the amplified gas radial velocity, \citet{2012ApJ...755....6Z} find a relation between the dust depletion factor ($\gamma$/$\gamma_{0}$) and the gaseous gap depth ($\Sigma_{g}/\Sigma_{g, 0}$), where $\gamma$ and $\Sigma_{g}$ is the  dust/gas mass ratio and the surface density inside the gap, respectively, and $\gamma_{0}$ and $\Sigma_{g, 0}$ are these quantities at the gap edge. The relationship between $\gamma$/$\gamma_{0}$ and $\Sigma_{g}/\Sigma_{g, 0}$ depends on the dust particle density ($\rho_{p}$), dust particle radius ($s$), gas surface density at the gap edge ($\Sigma_{g, 0}$), and disk viscosity parameter ($\alpha$). Following the analytic approach in \citet{2012ApJ...755....6Z}, we study the efficiency of dust filtration in the disk of GW~Ori using the disk parameters (Type~1 in Table~\ref{Tab:disk_par}) of a best-fit model shown in Fig.~\ref{Fig:SED}. We take $\rho_{p}$=2.5\,g/cm$^{-3}$ for the silicate grain, and $\alpha$=0.01. Figure~\ref{Fig:Dust}~(a) shows the relation between $\gamma$/$\gamma_{0}$ and $\Sigma_{g}/\Sigma_{g, 0}$ for dust particles with four sizes, $s$=0.1, 1, 10, and 100\,\mum. Note that a gaseous gap depth$\sim$0.04 and 0.004 can efficiently ($\gamma$/$\gamma_{0}$=$10^{-3}$) trap dust particles with sizes 100\,\mum\ and 10\,\mum, respectively. To trap smaller dust particles, a deeper gaseous gap is required.

\begin{figure}
\centering
\includegraphics[width=1.\columnwidth]{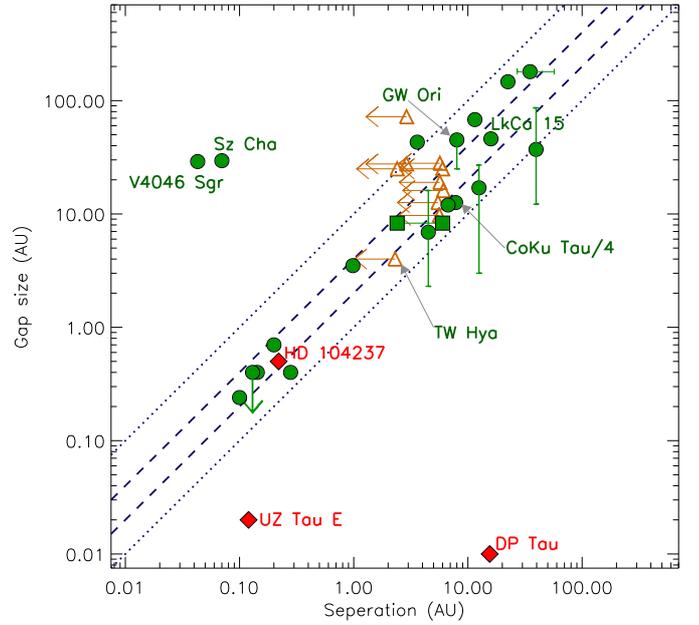}
\caption{Gap size of a disk surrounding a binary/multiple stellar system vs. separation from the companion to the primary. The filled circles show the disks with gaps, and the filled diamonds are for the disks without gaps. The filled boxes connected by a solid line are for the source FL~Cha of which  two possible separations are given in \citet{2013ApJ...762L..12C}. The open triangles show the disks with gaps but without known companions. The separations for these disks are the detection limits at given mass ratios of companions to primaries (see Appendix~\ref{Appen:binary}). The left-pointing arrows are used to indicate that it could be still possible that there are companions at smaller distance than the detection limits. The dotted lines mark the 1:1 and 10:1 relation between gap sizes and separations. The dashed lines show the predicted relation between gap sizes and separations from \citet{2008MNRAS.391..815P} with $e$=0 and 1 \label{Fig:Gap}}
\end{figure}

The dust drift velocity ($V_{\rm drift}$) needs to counteract not only the dust diffusion but also  the amplified gas radial velocity ($V_{\rm g}$) in the gap. In Fig.~\ref{Fig:Dust}(a), we do not consider the effect of amplified gas radial velocity in the gap. Thus, the dust depletion factors are only the lower limits for dust particles at the gaseous gap depths. When the effect of amplified gas radial velocity is included in the calculation, \citet{2012ApJ...755....6Z} built another relationship between ratios of $V_{\rm drift}$ to $V_{\rm g}$ and gaseous gap depths. In Fig.~\ref{Fig:Dust}~(b), we show ratios of $V_{\rm drift}$ to $V_{\rm g}$ with respect to gaseous gap depths for  dust particles with sizes of 0.1, 1, 10, and 100\,\mum. When $V_{\rm drift}$ is larger than $V_{\rm g}$, the dust particles are trapped. For a 10\,\mum-size dust grain, it can be trapped when $\Sigma_{g}/\Sigma_{g, 0}$$\sim$0.0003. At a lower $\Sigma_{g}/\Sigma_{g, 0}$, smaller dust particles can be trapped. Our SED modeling can only provide the dust surface density in the disk. If we assume the gas-to-dust ratio  is constant, $\Sigma_{g}/\Sigma_{g, 0}$ is $\sim$1.5$\times$10$^{-6}$ for the disk model in Fig.~\ref{Fig:SED}. In this case, dust particles larger than 10\,\mum\ can be efficiently filtered. However, it would be highly probable that the gas-to-dust ratio in the gap is higher than that in the outer disk if the dust filtration is occurring in the gap. Thus, the gaseous gap depth is expected to be more shallow than that from the SED modeling. Furthermore, our discussions here are tentative since they are based on the disk parameters from the SED modeling, which are very uncertain.

\subsubsection{Gravitational instability}

The strong continuum emissions at submillimeter and millimeter wavelength suggest that GW~Ori system is harboring a massive disk \citep{1995AJ....109.2655M}. Such a disk may suffer from the gravitational instability. A criterion to judge the gravitational instability in a disk is given by the Toomre parameter ($Q$) \citep{1964ApJ...139.1217T}:

\begin{center}
\begin{equation}
 Q=\frac{c_{\rm s}\Omega}{2\pi{\rm G}\Sigma},\nonumber
\end{equation}
\end{center}

\noindent where $c_{\rm s}$ is the sound speed, $\Omega$ is the Keplerian angular velocity, G is the gravitational constant,  and $\Sigma$ is the surface density of disk. The instability occurs in the disk when $Q<$1, in which the disk is sufficiently cool, or massive, and when $Q>$1, in which the disk is stable and not fragmented. We calculate the $Q$ value for the circumbinary disk of GW~Ori using the disk parameters (Type~1 in Table~\ref{Tab:disk_par}) for the model shown in Fig.~\ref{Fig:SED}. The Toomre parameter decreases with the disk radius but is always larger than 1, indicating that the disk of GW~Ori is stable at all radii. The other two types of disk models in Fig.~\ref{Fig:var_SED} and Table~\ref{Tab:disk_par} are only different from the type~1 model in the inner-disk structure ($<$45\,AU), and are also stable at all the disk radii.

\subsection{Gap sizes and binary separations}

The study toward several star-forming regions, including Taurus, Cha\,I, and Ophiuchus, reveals about 2/3 of close binaries ($\lesssim$40\,AU) have dispersed their disks at ages $<$1\,Myr, thus providing important constraints on the timescale of planet formation in such systems \citep{2012ApJ...745...19K}. A fast dissipation of disks in close binaries is expected from theoretical works \citep{1993prpl.conf..749L}. In the inner region of a circumbinary disk, a gap  quickly forms and isolates the circumstellar disks from the  circumbinary disk. Both disks  evolve independently. Based on the calculations of \citet{1994ApJ...421..651A}, a circumbinary disk can be truncated inside at  $R_{\rm gap}\sim1.8a-2.6a$, where $a$ is the orbital semi-major axis of a binary.  \citet{2008MNRAS.391..815P} find the similar result. They run the simulations with mass ratios $0.1\leq q\leq 0.9$ and eccentricity $0\leq e\leq 0.9$ and find  $R_{\rm gap}\approx 1.93a(1+1.01e^{0.32})[q(1-q)]^{0.043}$. For GW~Ori, the circumbinary disk is probably truncated by GW~Ori~C, which gives $R_{\rm gap}\sim$2$a$--4$a$ at extreme cases of $e$=0 and 1, by assuming $q>$0.1. Given a projected separation of $\sim$8\,AU between GW~Ori~A and GW~Ori~C, $R_{\rm gap}$ is around 16--32\,AU, which is comparable to the result (25--55\,AU) from our SED modeling.   

In Fig.~\ref{Fig:Gap}, we collect a sample of binary/multiple systems with circumbinary disks and show their gap sizes  with respect to the separations ($a$) between the companions and the primaries. The detail information for each binary is described in Appendix~\ref{Appen:binary}. Here, the separations shown in Fig.~\ref{Fig:Gap} are mostly projected values and thus present  lower limits of the real values. Among the sources in Fig.~\ref{Fig:Gap}, three systems, including UZ~Tau~E, DP~Tau, and HD\,104237, may be harboring circumbinary disks without any evidences of gaps. The systems UZ~Tau~E and DP~Tau could be too young to clear out their inner disks. The object HD\,104237 is A-type hot young stars. Its inner radius of a disk set by the dust evaporation is around 0.5\,AU, comparable to the one that are supposed to be truncated by the companion. In Fig.~\ref{Fig:Gap}, we also see that two systems, V4046~Sgr and Sz~Cha, which have very small separations ($<$0.1\,AU) but are harboring disks with much larger gaps than the expectation from the theory \citep{1994ApJ...421..651A,2008MNRAS.391..815P}. The accretion rates of these sources are typically lower \citep[2$\times$10$^{-9}$--6$\times$10$^{-10}$\,\accunit,][]{2011MNRAS.417.1747D,2011ApJ...728...49E}, and the photoevaportion process is expected to take effect in the disk dissipation and clear out the  inner regions of disks \citep{2006MNRAS.369..229A,2011MNRAS.412...13O}. However, it cannot be excluded that in these systems there are  ``unseen'' companions at  separations  large enough to create the big gaps in the disks.

To test the significance of the correlation between the gap sizes and separations in Fig.~\ref{Fig:Gap}, we apply a Kendall $\tau$ test. The Kendall $\tau$ test yields $\tau$=0.52 and $p$=3$\times$10$^{-4}$ for all the sources in Fig.~\ref{Fig:Gap}, suggesting the gap sizes and separations are clearly correlated. If we exclude the ``outliers'' discussed in the above paragraph, the correlation between the values are more significant with $\tau$=0.81 and $p$=1$\times$10$^{-6}$ from the  Kendall $\tau$ test. In Fig.~\ref{Fig:Gap}, we also show the relation between the gap sizes and the separations  predicted by \citet{2008MNRAS.391..815P} for $e$=0 and 1. We note the sources in  Fig.~\ref{Fig:Gap} follow the relations in general but are distributed within a broad ``band'' (a large scatter). This can be due to the large uncertainties in both the gap sizes and the separations, since the separations are mostly projected ones and the gap sizes are usually estimated from SED modeling (see Appendix~\ref{Appen:binary}). In Fig.~\ref{Fig:Gap}, about 56\% of systems have a value of $1<R_{\rm gap}/a<4$ and the median value of $R_{\rm gap}/a$ is around 3, which is consistent with the theoretical prediction \citep{1994ApJ...421..651A,2008MNRAS.391..815P}.

In Fig.~\ref{Fig:Gap}, we also show a collection of transition objects (TOs). These sources have been observed with high-resolution imaging and do not show any stellar or substellar companions at separations less than several AUs, although they are known to have gaps in the disks. {\rev We use the detection limits of these observations as the upper limits of separations between ``unseen'' companions and the central stars.} For these TOs, some other physical mechanisms, such as photoevaportion, may be responsible for the clearing of their inner disks. However, we note that most of these TOs are still located within the broad ``band'', where the binary systems are distributed in the figure. It could be still possible that there are ``unseen'' companions at  separations smaller than the detection limits of the observations.

\section{Summary}\label{Sec:summary}

We have studied the triple system GW~Ori with a special focus on the accretion properties and disk properties, using a combination of available archive data and our own high-resolution spectroscopic data from FEROS and HARPS as observed during 2007 and 2010. We summarize our main results as follows:

\begin{itemize}
\item By comparing the FEROS spectra with the synthetical spectra, we classify GW~Ori~A as G8 type. {\newrev The mass and age of GW~Ori is estimated to be $\sim$3.9\Msun and is less than 1\,Myr, which points out that GW~Ori~A could be at a rapid transitional phase to a Herbig~Be star.}

\item We analyze the RVs of the spectroscopic data of GW~Ori and confirm a companion around GW~Ori with a period of $\sim$242\,days and an orbital semi-major axis of $\sim$1\,AU. The residual RVs from the orbital solution and the EWs of accretion-related emission lines show periodic variations during short terms (5--6.7\,days). We explain that they are from the modulation with stellar rotation of GW~Ori~A.

\item {\newrev The variance profiles of the H$\alpha$ and H$\beta$ lines of GW~Ori present blue-shifted peaks and are featureless on the red side. The autocorrelation matrices for the  H$\alpha$ and H$\beta$ lines show a clear correlation between the blue ($-300$--0\,\kms) and the red (0--200\,\kms) sides of profiles. The H$\alpha$ and H$\beta$ line profiles can be decomposed very well by two emission components and one blue-shifted absorption component.}  The absorption component can be attributed to the disk wind launched at a disk radius near the orbit of GW~Ori~B. The strength of the disk wind can be modulated by the orbital motion of GW~Ori~B. 

\item We investigate the accretion behavior of GW~Ori using the accretion-related emission lines and $U$-band photometry. Our results contradict the simulations of \citet{1996ApJ...467L..77A} at two points: (1) the accreted material of GW~Ori appears to be mostly funneled onto the primary GW~Ori~A; (2) the accretion rates of GW~Ori are mostly constant and are only occasionally enhanced  by a factor of 2--3 within the orbital phases of 0.5--0.8. {\subrev At Point (1), our result is consistent with the new simulations from \citet{2011Ap&SS.335..125F}.} 

\item We reproduce the SED of GW~Ori using disk models with gaps sized 25--55\,AU. The gap sizes in our models are much larger than the previous result (3.3\,AU) in \citet{1995AJ....109.2655M}, {\subrev which are comparable to the truncation of the inner disk radius (16--32\,AU), as expected from the simulations, considering the second companion GW~Ori~C.} In the gap, only a small amount of tiny dust particles are added to produce the near-infrared excess emission and the strong and sharp silicate emission feature at 10\,\mum. Using the best-fit parameters of the disk models, we study the efficiency of dust filtration in the disk of GW~Ori and find that dust grains larger than 10\,\mum\ can be efficiently trapped in the outer disk.

 \item We find that GW~Ori {\subrev shows dramatic changes in its SED in near-infrared bands, which implies a major readjustment of the inner disk on timescales of $\sim$20\,yr.} We discuss the possibility that the material in the gap is drained by the accretion in GW~Ori and find that the timescale to exhaust the material in the gap  via accretion can be several years only if the gas-to-dust ratio in the gap is $\sim$100-1000, which can  explain the rapid variation in the SED of GW~Ori in near-infrared bands.

\item A sample of binary/multiple systems collected in the literature exhibits a strong positive correlation between their gap sizes and the separations of companions to the primaries, which is consistent with the expectation from the theory.

\end{itemize}

\begin{table*}
\caption{Parameters for young systems with companions/planets.\label{Tab:allbinary}}
\centering
\begin{tabular}{cccccccc}
\hline
\multicolumn{7}{c}{{\bf With Gap}}\\
\hline
                   &                 &Separation            &           &  & Gap Size      &        &  \\
Name               &Mass ratio       &(AU)         &Method1 &Eccentricity          & (AU)     &Method2  &Reference \\  
\hline
CoKu~Tau/4         &0.85             &7.8          &Image    &\nodata     & 12.6      &SED Fit    &1, 2\\
T~Cha              &\nodata          &6.7          &Image    &\nodata     & 12        &Mid-IR Int &3, 4\\ 
V4046~Sgr          &0.96             &0.04$^{a}$    &Spec, D &$<$0.01     & 29        &MM Int     &5, 6\\
AK~Sco             &0.987            &0.14         &Spec, D  &0.47        &0.4        &SED Fit    &7\\
GG~Tau~A           &0.9              &35$^{+22}_{-8}$$^{b}$&Image   &0.3$\pm$0.2 &180     &MM Int     &8, 9, 10\\
CS~Cha             &\nodata          &$\sim$3.6$^{c}$ &Spec, S  &\nodata     &43         &SED Fit    &11, 12, 13\\ 
HD~98800~B          &0.83             &0.98        & Spec, D  &0.78        &3.5        &SED Fit    &14, 15, 16\\
DQ~Tau             &0.97             &0.13         &Spec, D   &0.556       &$\lesssim$0.4 &SED Fit    &17, 18\\
162814--2427       &0.92             &0.28         &Spec, D  &0.48        &0.4        &SED Fit    &19, 20\\ 
162819--2423S      &$\gtrsim$0.26    &0.1          &Spec, S  &0.41        &0.24       &SED Fit    &19, 20\\
FL~Cha             &\nodata          &2.4/6        &Image    &\nodata     &8.3        &SED Fit    &21\\ 
LkH$\alpha$~330    &\nodata          &10--13       &Image    &\nodata     &68         &MM Int     &22, 23\\
T21                &\nodata          &22.4         &Image    &\nodata     &146.7      &SED Fit     &24, 25\\
DF~Tau             &0.75             &12.5$^{d}$    &Image    &0.51        &17$^{+10}_{-14}$  &SED Fit &26, 27, 28\\
MHO~3              &0.5              &4.5          &Image    &\nodata     &6.9$^{+9.2}_{-4.6}$ &SED Fit &29, 30\\
CHX22              &\nodata          &39.5         &Image    &\nodata     &37.1$^{+49.3}_{-24.9}$ &SED Fit &24, 30\\
ST~34              &$\sim$1          &$\sim$0.2$^{e}$&Spec, D &\nodata     &0.7        &SED Fit  &31, 32\\
Sz~Cha             &                 &$\sim$0.07$^{f}$&Spec, S  &\nodata     &29.5       &SED Fit  &33, 30 \\
LkCa~15            &$\sim$0.006      &$\sim$15.9   &Image    &\nodata     &50         &MM Int     &34, 35 \\
\hline
\multicolumn{7}{c}{{\bf  Without Gap}}\\
\hline
                   &                 &Separation         &    &             & $R_{\rm in}$      &        &  \\
Name               &Mass ratio       &(AU)        &Method1   &Eccentricity          & (AU)     &Method2  &Reference \\  
\hline
%HD~104237          &0.64              &0.22$^{h}$ &Spec, S   &0.66         &0.5        &SED Fit  &38, 39, 40 \\    
HD~104237          &0.64              &0.22$^{g}$ &Spec, S   &0.66         &0.5        &SED Fit  &36, 37, 38 \\    
%UZ~Tau~E           &0.30              &0.124      &Spec, D   &0.33         &$\sim$0.02  &$R_{\rm sub}$    &41, 42\\
UZ~Tau~E           &0.30              &0.124      &Spec, D   &0.33         &$\sim$0.02  &$R_{\rm sub}$    &39, 40\\
%DP~Tau             &0.74              &15.5       &Image     &\nodata      &$\sim$0.01  &$R_{\rm sub}$    &29, 43\\
DP~Tau             &0.74              &15.5       &Image     &\nodata      &$\sim$0.01  &$R_{\rm sub}$    &29, 41\\
 \hline
\multicolumn{7}{c}{{\bf TOs without known companions and their detection limit}}\\
\hline
                   &                 &Separation            &            &             & Gap Size      &        &  \\
Name               &Mass ratio       &(Detection limit, AU)    &Method1     &Eccentricity          & (AU)     &Method2  &Reference \\  
\hline

GM~Aur              &      0.01      &5.7                  &Image       &\nodata         &28        &MM Int  &29, 22\\    
UX~Tau~A            &      0.01      &6           &Image    &\nodata    &25        &MM Int  &29, 22\\ 
RY~Tau              &      0.01      &2.9          &Image   &\nodata     &27.6      &SED Fit &29, 30\\ 
DM~Tau              &      0.02      &5.7         &Image     &\nodata   &19        &MM Int  &29, 22\\    
GK~Tau              &      0.02      &5.6          &Image   &\nodata    &9.7       &SED Fit &29, 22\\ 
HK~Tau              &      0.02      &6.1         &Image    &\nodata    &16.1      &SED Fit &29, 30\\ 
%RX~J1604.3--2130     &      0.02      &3          &Image    &\nodata    &72        &MM Int  &44, 45\\    
RX~J1604.3--2130     &      0.02      &3          &Image    &\nodata    &72        &MM Int  &42, 43\\    
IP~Tau              &      0.03      &5.5         &Image    &\nodata    &12.6      &SED Fit &29, 30\\ 
%RX~J1633.9--2442    &$\sim$0.02      &2.4         &Image     &\nodata   &$\sim$25  &MMT Int &46\\ 
RX~J1633.9--2442    &$\sim$0.02      &2.4         &Image     &\nodata   &$\sim$25  &MMT Int &44\\ 
TW~Hya        &\nodata         &2.3         &Image     &\nodata    &4          &SED Fit     &45, 46 \\

\hline
%\multicolumn{7}{c}{{\bf Binary systems with outer truncated disks}}\\
%\hline
%                   &                 &Seperation              &         &             & $R_{\rm out}$      &        &  \\
%Name               &Mass ratio       &(AU)             &Method1        &Eccentricity & (AU)     &Method2  &Reference \\  
%\hline
%ZZ~Tau             &0.48             &6.1              &Image          &\nodata      &$\sim$3          &SED Fit   &29, 47 \\
%SR~20              &\nodata          &5.3--9.9         &Image          &\nodata      &0.39             &SED Fit   &48, 49, 50 \\
%HD~98800~AB        &0.68             &38               &Image          &\nodata      &10--15           &MM Int    &51, 16 \\
%Hen~3--600AB       &\nodata          &66               &Image          &\nodata      &15--25           &MM Int    &52, 16 \\
%L1551              &                 &
%\hline
\end{tabular}
\tablebib{1. \citet{2008ApJ...678L..59I}; 2. \citet{2010ApJ...708...38N}; 3. \citet{2011A&A...528L...7H}; 4. \citet{2013A&A...552A...4O}; 5. \citet{2004A&A...421.1159S}; 6. \citet{2013ApJ...775..136R}; 7. \citet{2003A&A...409.1037A}; 8. \citet{1999ApJ...520..811W}; 9. \citet{2002ApJ...575..974M}; 10. \citet{1999A&A...348..570G}; 11. \citet{2007A&A...467.1147G}; 12. \citet{2007ApJ...664L.111E}; 13. \citet{2012ApJ...747..139N}; 14. \citet{1995ApJ...452..870T}; 15. \citet{2005ApJ...635..442B}; 16. \citet{2010ApJ...710..462A}; 17. \citet{1997AJ....113.1841M}; 18. \citet{2009ApJ...696L.111B}; 19. \citet{1989AJ.....98..987M}; 20. \citet{1997AJ....114..301J}; 21. \citet{2013ApJ...762L..12C}; 22. \citet{2011ApJ...732...42A}; 23. \citet{2013ApJ...775...30I}; 24. \citet{2008ApJ...683..844L}; 25. \citet{2009ApJ...700.1017K}; 26. \citet{2001ApJ...556..265W}; 27. \citet{1993ApJ...405L..71M}; 28. \citet{2002ApJ...578..925T}; 29. \citet{2011ApJ...731....8K} 30. \citet{2013ApJ...769..149K}; 31. \citet{2005ApJ...621L..65W}; 32. \citet{2005ApJ...628L.147H}; 33. \citet{2002AJ....124.2813R}; 34. \citet{2012ApJ...745....5K}; 35. \citet{2006A&A...460L..43P};  36. \citet{2004A&A...427..907B}; 37. \citet{2007A&A...464...55T}; 38. \citet{2013MNRAS.430.1839G}; 39. \citet{2007AJ....134..241J}; 40. \citet{1996AJ....111.2431J}; 41. \citet{2006ApJS..165..568F}; 42. \citet{2008ApJ...679..762K}; 43. \citet{2012ApJ...753...59M}; 44. \citet{2012ApJ...752...75C}; 45. \citet{2002ApJ...568.1008C}; 46. \citet{2003AJ....126.2009B}.} %; 47. \citet{2012ApJ...747..103E}; 48. \citet{1993AJ....106.2005G}; 49. \citet{1995AJ....110..753G}; 50. \citet{2008ApJ...683L.187M}; 51.\citet{2001ApJ...549..590P}; 52. \citet{1989ApJ...343L..61D}}
\tablefoot{Column 3: If it is not specified,  the separation is projected distance between two stellar components when the binary is detected with imaging, and is minimum semi-major of the orbit when it is detected with spectroscopy. Column 4: The methods with which the binaries are detected: Image for imaging, Spec for spectroscopy, D for double lines, and S for single line. Column 4: the methods with which the gap sizes are estimated, including fitting the SED (SED Fit), mid-infrared interferometry (Mid-IR Int), millimeter interferometry (MM Int). For the full disks (UZ\,Tau\,E and DP\,Tau), the inner radius of the disks are determined by the the evaporation of dust ($R_{\rm sub}$). a: sum of semi-majors of the orbits for the two stellar components with an assumption of  inclination 35$^{\circ}$. b, d: the orbital semi-major from fitting the orbit. c: roughly estimated using the orbit period 2482\,day. e: roughly estimated assuming the system is edge-on. f: roughly estimated using the orbit period of 5\,days.g: the semi-major of the orbit  assuming an inclination 17$^{\circ}$.}
\end{table*}
\normalsize

\begin{acknowledgements} 
We thank  the referee, Dr.~S.~Lamzin, for the positive and useful comments. We want to acknowledge C. Dullemond for providing us with his RADMC code. MF acknowledges support of the action ``Proyectos de Investigaci\'{o}n fundamental no orientada", grant number AYA2012-35008. ASA support of the Spanish MICINN/MINECO ``Ram\'on y Cajal'' program, grant number RYC-2010-06164, and the action ``Proyectos de Investigaci\'on fundamental no orientada'', grant number AYA2012-35008. V.R. was partially supported by the {\it Bayerischen Gleichstellungsf{\"o}rderung (BGF)}. This research has made use of the SIMBAD database, operated at CDS, Strasbourg, France. This publication makes use of data products from the Two Micron All Sky Survey, which is a joint project of the University of Massachusetts and the Infrared Processing and Analysis Center/California Institute of Technology, funded by the National Aeronautics and Space Administration and the National Science Foundation. This publication makes use of data products from the Wide-field Infrared Survey Explorer, which is a joint project of the University of California, Los Angeles, and the Jet Propulsion Laboratory/California Institute of Technology, funded by the National Aeronautics and Space Administration. This research is based on observations with AKARI, a JAXA project with the participation of ESA. This work is in part  based  on observations made with the Spitzer Space Telescope, which is operated by the Jet Propulsion Laboratory, California Institute of Technology under a contract with NASA. 
\end{acknowledgements} 

\begin{appendix}
\section{Young binary systems}\label{Appen:binary}
We collect a sample of binary systems with disks and present them in Table~\ref{Tab:allbinary}. We list the companion/primary mass ratios, the separations, then eccentricities of the binaries, the gap sizes of the disks, and the methods using which the gap sizes are derived. In the table, the young star LkCa~15 with a possible planet is also listed. As a comparison, we also list three spectroscopic binaries without sign of gaps in their disks in Table~\ref{Tab:allbinary}.

In Table~\ref{Tab:allbinary}, ten TOs  are listed. \citet{2008ApJ...679..762K} and \citet{2011ApJ...731....8K,2012ApJ...752...75C} have imaged these TOs with a high spatial resolution and found no companions. The separations of these TOs are on a lower limit of distances from the central stars, where the observations can detect companions at given companion/primary mass ratios.

\end{appendix}

\bibliographystyle{aa}
\bibliography{references}

\end{document}